\documentclass[sigconf]{acmart}
\setcopyright{acmlicensed}
\copyrightyear{2026}
\acmYear{2026}
\acmDOI{XXXXXXX.xxxxxXX}
\setcopyright{none}
\renewcommand\footnotetextcopyrightpermission[1]{}
\acmISBN{978-1-4503-XXXX-X/2026/06}

\usepackage[utf8]{inputenc}
\usepackage[most]{tcolorbox}
\usepackage{xspace}
\usepackage{xcolor}      
\usepackage{listings}    
\usepackage[english]{babel} 
\usepackage{microtype}      
\usepackage{underscore}     
\usepackage{enumitem}

\usepackage[T1]{fontenc}
\usepackage[utf8]{inputenc}
\usepackage[utf8]{inputenc}
\usepackage{tikz}
\usepackage{array} 
\usepackage{graphicx}
\usepackage{tabularx}
\usepackage{multirow}
\usepackage{threeparttable}

\usepackage{enumitem}
\usepackage{algorithmic}
\usepackage{amsmath}
\usepackage{booktabs}
\usepackage{geometry}
\usepackage{subcaption}
\usepackage[ruled,vlined,linesnumbered]{algorithm2e}
\usepackage{tikz}

\newcommand{\blackcircled}[1]{%
  \tikz[baseline=(char.base)]{%
    \node[shape=circle, fill=black, text=white, minimum size=0.8em, inner sep=0] (char) {#1};%
  }%
}

\newtheorem*{definition*}{Definition}

\definecolor{myred}{RGB}{255,0,0}

\newcommand{\tool}{\textsc{Clover}\xspace}
\newcommand{\racebench}{RaceBench 2.1\xspace}
\newcommand{\purellm}{DRB-LLM\xspace}
\newcommand{\code}[1]{\texttt{#1}}

\definecolor{todocolor}{rgb}{0.9,0.1,0.1}

\newcommand{\eg}{\hbox{\emph{e.g.}}\xspace}
\newcommand{\ie}{\hbox{\emph{i.e.}}\xspace}

\definecolor{codegreen}{rgb}{0,0.6,0}
\definecolor{codegray}{rgb}{0.5,0.5,0.5}
\definecolor{codepurple}{rgb}{0.58,0,0.82}
\definecolor{backcolour}{rgb}{0.97,0.97,0.95}
\definecolor{forestgreen}{rgb}{0.28,0.62,0.37}

\lstdefinestyle{mystyle}{
    backgroundcolor=\color{backcolour},   
    commentstyle=\color{codegray},
    keywordstyle=\color{codepurple},
    numberstyle=\tiny\color{codegray},
    stringstyle=\color{blue},
    basicstyle=\ttfamily\footnotesize,
    breakatwhitespace=false,         
    breaklines=true,                 
    captionpos=b,                    
    keepspaces=true,                 
    numbers=left,                    
    numbersep=5pt,                  
    showspaces=false,                
    showstringspaces=false,
    showtabs=false,                  
    tabsize=4,
}

\begin{document}
\title{Automated detection of atomicity violations in large-scale systems}
\author{Hang He}
\authornote{These authors contributed equally to this work.}
\affiliation{%
  \institution{East China Normal University}
  \city{Shanghai}
  \country{China}}
\email{hang.he@stu.ecnu.edu.cn}

\author{Yixing Luo}
\authornotemark[1]
\affiliation{%
  \institution{Beijing Institute of Control Engineering}
  \city{Beijing}
  \country{China}}
\email{luoyi_xing@126.com}

\author{Chengcheng Wan}
\authornote{Chengcheng Wan and Ting Su are corresponding authors.}
\affiliation{%
  \institution{East China Normal University
  Shanghai Innovation Institute}
  \city{Shanghai}
  \country{China}}
\email{ccwan@sei.ecnu.edu.cn}

\author{Ting Su}
\authornotemark[2]
\affiliation{%
  \institution{East China Normal University}
  \city{Shanghai}
  \country{China}}
\email{tsu@sei.ecnu.edu.cn}

\author{Haiying Sun}
\affiliation{%
  \institution{East China Normal University}
  \city{Shanghai}
  \country{China}}
\email{hysun@sei.ecnu.edu.cn}

\author{Geguang Pu}
\affiliation{%
  \institution{East China Normal University}
  \city{Shanghai}
  \country{China}}
\email{ggpu@sei.ecnu.edu.cn}

\begin{abstract}
Atomicity violations in interrupt-driven programs pose a significant threat to software reliability in safety-critical systems. These violations occur when the execution sequence of operations on shared resources is disrupted by asynchronous interrupts. Detecting atomicity violations is challenging due to the vast program state space, application-level code dependencies, and complex domain-specific knowledge. In this paper, we propose \tool, a multi-agent framework for detecting atomicity violations in real-world interrupt-driven programs. Its plan agent orchestrates four static analysis tools to extract key information and generate code summaries. \tool then initializes several \emph{Expert-Judge} agent pairs to detect and validate different patterns of atomicity violation, through an iterative manner.
Evaluations on \racebench, SV-COMP, and RWIP demonstrate that \tool achieves a precision/recall of 91.0\%/96.4\%, outperforming existing approaches by 33.0-117.2\% on F1-score.

\end{abstract}



\maketitle

\section{Introduction}
\sloppy
\subsection{Motivation}
\label{sec.1.1}

Interrupt-driven programs are integral components of modern safety-critical systems that require timely and reliable responses to external events, such as aerospace, automotive electronics, and medical equipment~\cite{8115634, 6004502,parmer2008predictable,leveson1983analyzing,lu2011finding}. These programs leverage \emph{Interrupt Service Routines} (ISRs) to handle asynchronous events, which introduces non-deterministic instruction execution orders~\cite{jin2011automated,7318260,regehr2005random}. As a result, developers often struggle to avoid atomicity violation (AV) bugs, which occur when the operation sequence on shared resources (\eg, variables) is disrupted by concurrent ISR execution~\cite{park2009ctrigger,gu2015change,halstead1985multilisp}. Such violations undermine software safety and correctness, underscoring the need for effective detection solutions~\cite{lu2008learning,7092387,pradel2014performance,tu2019understanding}.

\begin{figure}[]
\centering
\includegraphics[width=\linewidth]{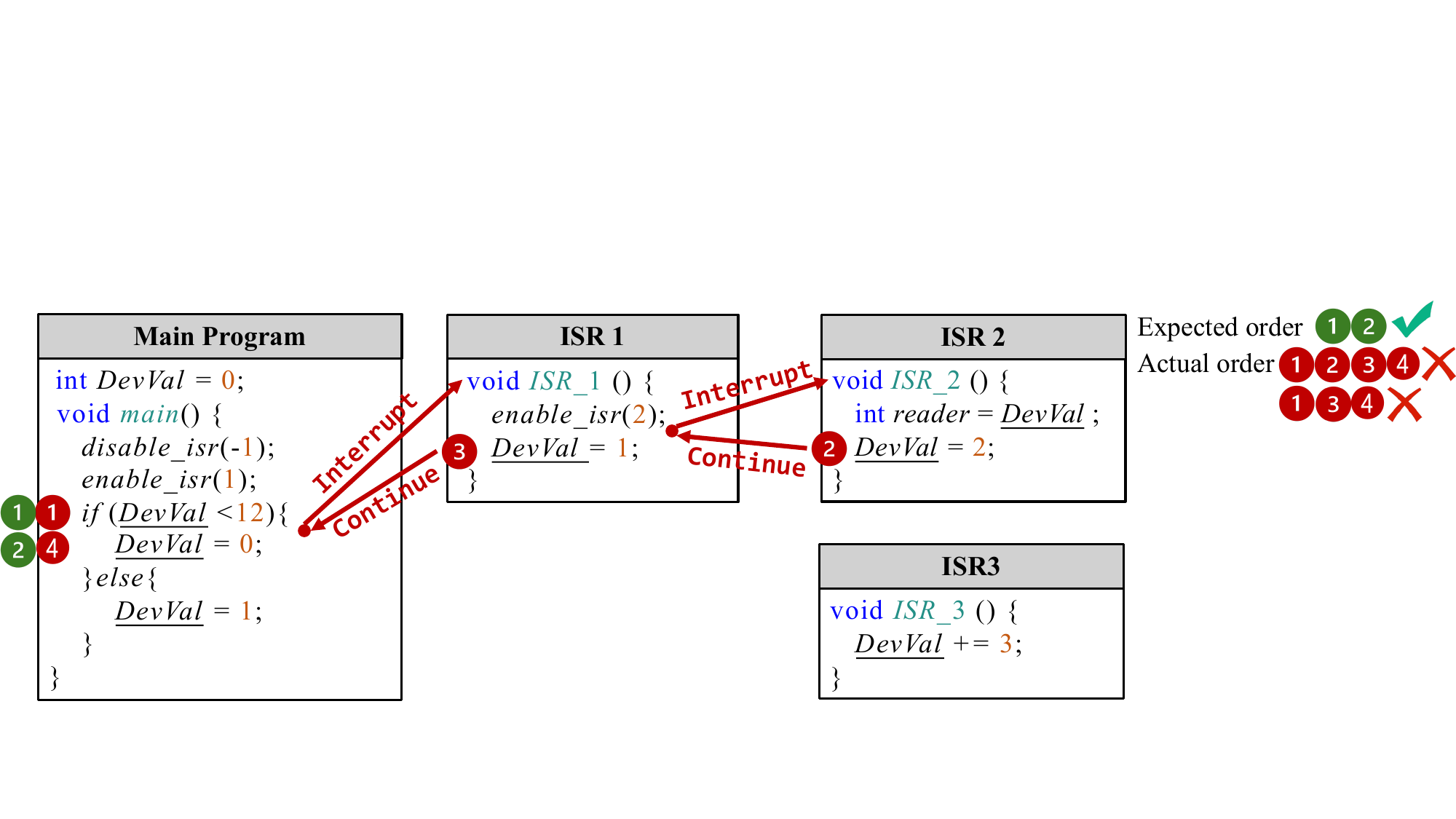}
\caption{An atomicity violation example from \racebench~\cite{racebench}. 
}
\label{fig:motivation}
\end{figure}

To illustrate atomicity violation bugs, consider a simplified example from RaceBench 2.1~\cite{racebench}. As shown in Figure~\ref{fig:motivation}, the global shared variable \texttt{DevVal} is accessed by both the main program (\emph{main}) and three ISRs, with \emph{ISR\_}3 having the highest priority, followed by \emph{ISR\_}2, \emph{ISR\_}1, and \emph{main} the lowest. Initially, all ISRs are disabled by \texttt{disable\_isr(-1)}. The main program then enables \emph{ISR\_}1, which in turn enables \emph{ISR\_}2; \emph{ISR\_}3 remains disabled throughout execution. 
The main program is expected to execute read and write operations on \texttt{DevVal} atomically, as indicated by the green circles (\blackcircled{1}-\blackcircled{2}) in the figure, ensuring the non-fall-through edge is executed when the branch condition is satisfied. However, due to interrupt preemption, the main program may be interrupted by \emph{ISR\_}1 between these two operations \blackcircled{1} and \blackcircled{2}, and \emph{ISR\_}1 is further preempted by \emph{ISR\_}2 before writing to \texttt{DevVal}. As shown in the red circles, problematic interleavings (\blackcircled{1}-\blackcircled{2}-\blackcircled{3}-\blackcircled{4} and \blackcircled{1}-\blackcircled{3}-\blackcircled{4}) breaks the atomicity on \texttt{DevVal} in the main program, making the write operation to overwrite modifications made by the ISRs. Thus, the conditional logic in the main program is inconsistent with its actual action.

This type of defect turns out to be hard to detect. Interrupts can be triggered at unpredictable
times, leading to a huge state space in interrupt-driven programs~\cite{lu2007muvi,li2022precise,regehr2005random,mogul1997eliminating}. It is computationally heavy to exhaustively explore all possible instruction execution orders to identify potential atomicity violations. Although only containing 18 lines of code (\emph{ISR\_}3 excluded), Figure~\ref{fig:motivation} actually involves five states caused by different ISR trigger timings. Furthermore, shared resources are often referenced through intricate mechanisms, including function call chains and conditional branches. This further complicates the analysis of execution traces.

This example has demonstrated several open challenges in the automated detection of atomicity violations in interrupt-driven programs.

\textbf{1) Program state space reasoning.} 
An atomicity violation occurs only when a shared resource is operated in an improper sequence, which requires complex reasoning about operation sequences and atomicity. 

Several studies~\cite{lu2006avio, 10.1145/3597926.3598140, feng2020rchecker, 8812085, 6004502, yu2023detecting, li2022precise,10.1145/3695988} have proposed static analysis techniques to detect atomicity violations. However, they face the exhaustive exploration of the state space that grows exponentially with the possible preemption scenarios
and program scales. As a common drawback of static analysis approaches, they also struggle with complex code constructs such as pointers, dynamic data types, and intricate control flows, limiting their effectiveness in real-world applications. Recently, large language models (LLMs) and their agents have demonstrated promising capabilities in code comprehension and bug detection across various domains~\cite{zhao2024enhancing,hou2024large,tang2024codeagent,yildiz2025benchmarking,xu2022towards}. However, they focus shallow code anti-patterns and lack the reasoning capabilities required for temporal dependencies and concurrency semantics in interrupt-driven programs. 

\textbf{2) Application-level code context.} 
Not all program states are reachable during execution. To avoid false positives, it is essential to analyze all the relevant ISRs to identify reachable states. This requires application-level code analysis, which is a non-trivial task.

Traditional static analysis tools suffer from high computational overhead caused by state space explosions~\cite{cai2021canary,musuvathi2002cmc}. The non-determinism of interrupts leads to an exponential increase in the number of states as the program size grows~\cite{2022-20908}. Therefore, static analysis faces efficiency problems when applied to application-level code context. Similarly, LLMs encounter context limitations due to their fixed-size input windows~\cite{zeng2024memorize,huang2024recurrent}. They cannot capture long-range dependencies and program behavior that spans broader code sections, which is common in interrupt-driven programs.

\textbf{3) Complex domain-specific knowledge.} 
Atomicity violations often arise from complex interactions between hardware interrupts, synchronization mechanisms, and shared resources, requiring deep domain-specific knowledge to interpret. 

While static analysis tools are effective in understanding the timing logic of atomicity violations, they struggle to interpret program semantics, especially for third-party libraries or application scenarios. Meanwhile, LLMs are able to comprehend program semantics from identifier names~\cite{wang2023does,lehtinen2024let}. However, as pre-trained on general code corpus, they may lack the knowledge of interrupt-driven programs and timing constraints, which is critical to identifying atomicity violations.

\subsection{Contribution}

In this paper, we propose \tool, a novel end-to-end framework that integrates static analysis with LLM agents to detect atomicity violations in real-world interrupt-driven programs. To the best of our knowledge, this is the first work to leverage LLMs for atomicity violation detection.

To tackle the code context challenge, we design a suite of static analysis tools for different code semantics of interrupt-driven programs, and a \emph{Plan Agent} that dynamically orchestrates these tools through a unified framework. 
\emph{Plan Agent} first invokes \emph{Operation Analyzer} to identify shared resources between ISRs and the related operations.
For each shared resource, it then applies \emph{Defect Highlighter} to capture the potential defect locations for further examination. If the program to be analyzed exceeds the limit of LLM context window, a \emph{Code Extractor} is invoked to generate a compressed code snippet containing all the relevant code. It dynamically decides whether to invoke a \emph{Control Flow Analyzer} for inter-function control flow analysis. Therefore, the computational cost scales linearly with the size of the code.

To tackle the state space reasoning challenge, \tool proposes a hybrid reasoning mechanism that synergizes the precision of static analysis with the generalization capabilities of LLMs. For each shared resource, the static analysis tools precisely obtain its state information, including operation types and access patterns. Using this information, the LLM agents perform semantic reasoning on the compressed code snippet to determine whether any sequence of operations could lead to atomicity violations. This enables accurate and efficient analysis of real-world programs.

To tackle the domain-specific knowledge challenge, we design a knowledge-driven multi-agent system with (1) an \emph{Expert Agent} equipped with four knowledge modules, each dedicated to a specific type of atomicity violation; and (2) a \emph{Judge Agent} equipped with knowledge of execution contexts and interrupt interactions, examining whether a reported atomicity violation could actually be triggered. During the detection process, \tool invokes them iteratively to detect, validate, and refine bug reports, through a efficient to \emph{Conversation Manager}. Therefore, the defect hypotheses of \emph{Expert Agent} informs the validation focus of \emph{Judge Agent}, and the contextual assessments of \emph{Judge Agent} guide the \emph{Expert Agent} to refine its output, forming a complementary feedback loop.

We evaluate \tool against two static analysis approaches CPA4AV~\cite{yu2023detecting} and intAtom~\cite{li2022precise}, as well as an LLM-based approach DRB-LLM~\cite{guo2024largelanguagemodelbased}, on \racebench~\cite{racebench}, {SV-COMP}~\cite{SVCOMP2022} and RWIP~\cite{interruptdriven} dataset. The evaluation results clearly show the effectiveness of \tool: it achieves a precision/recall of  91.0\%/96.4\%, outperforming existing approaches by 33.0-117.2\% on F1-score.

\textbf{Data Availability.} To support the implementation of \tool, we release the detailed experimental results at~\url{https://anonymous.4open.science/r/clover-6EF6/}.

\section{Background}
\label{sec:section2}

\begin{figure}
    \centering 
    \includegraphics[width=\linewidth]{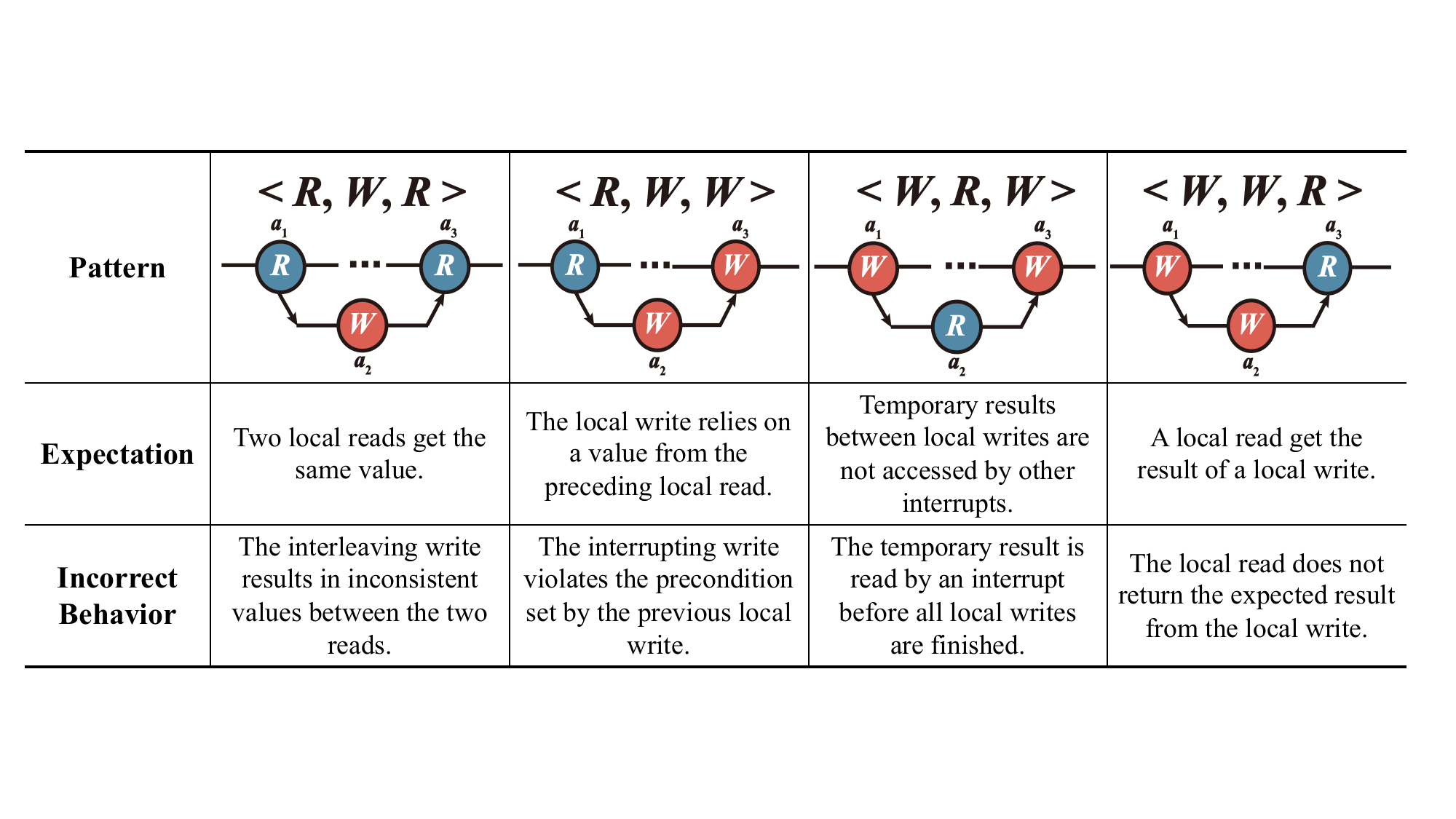}
    \caption{Four Atomic Violation Patterns. 
    }
    \label{fig:pattern}
\end{figure}

\subsection{Interrupt-Driven Program}
Interrupt-driven programs leverage interrupt mechanisms to manage concurrency in embedded systems, enabling real-time features~\cite{10.1145/3597926.3598140,8812085}. 
An interrupt-driven program typically contains one main program and several interrupt service routines (ISRs)~\cite{pan2019easy}. When an interrupt occurs, the corresponding ISR may or may not preempt the current execution flow, depending on its priority~\cite{schwarz2011static}. The communication between the main program and ISRs is realized by global shared variables.
Due to the non-deterministic nature of interrupts, the execution order in interrupt-driven programs is inherently unpredictable.

Each ISR is assigned a unique priority level. A high-priority ISR can preempt the execution of a low-priority ISR. Once the high-priority ISR is completed, the interrupted ISR resumes its execution. 
The main program is designated the lowest priority to ensure the preemption of all ISRs.

\subsection{Atomicity Violation}
In interrupt-driven programs, an atomicity violation occurs when the intended operation sequence of a global shared variable is disrupted by ISRs, violating the expected atomic order~\cite{10.1145/3597926.3598140,hofer2009sloth}. 
An atomicity violation is represented by a triple $<a_1, a_2, a_3>$, where each $a_i$ denotes a read or write operation on the same global shared variable~\cite{jin2011automated,farzan2009meta,lu2006avio}. A low-priority ISR is expected to execute $a_1$ and $a_3$ sequentially, while a high-priority ISR interleaves $a_2$ between them, violating atomicity.

\begin{definition*}
An atomicity violation occurs if there exists a low priority task $T_l$ and a high priority task $T_h$ such that:
\begin{itemize}
    \item $T_l$ contains two consecutive operations $a_1, a_3$ on a global shared variable $v$; and
    \item $T_h$ contains an operation $a_2$ on $v$; and
    \item $a_1, a_2, a_3$ contains at least one read operations; and
    \item The sequence of $<a_1, a_2, a_3>$ does not have read-after-read operations.
\end{itemize}
\end{definition*}

As shown in Figure~\ref{fig:pattern}, there are four atomicity violation patterns~\cite{lu2006avio}. The circles on upper lines are operations from low-priority ISRs, and the circles on lower lines are preempted operations from high-priority ISRs. All operations are performed on the same global shared variable. 

In the example of Figure~\ref{fig:motivation}, the global shared variable \texttt{DevVal} undergoes two Read-Write-Write sequences $<R, W, W>$, where the first write operation with the highest priority (from ISRs) is overwritten by the second one (from the main program) and thus violates atomicity.


\section{\tool: Atomicity Violation Detection}
\label{sec:section3}
\subsection{Overview}

\begin{figure*}
    \centering 
    \includegraphics[width=1\linewidth]{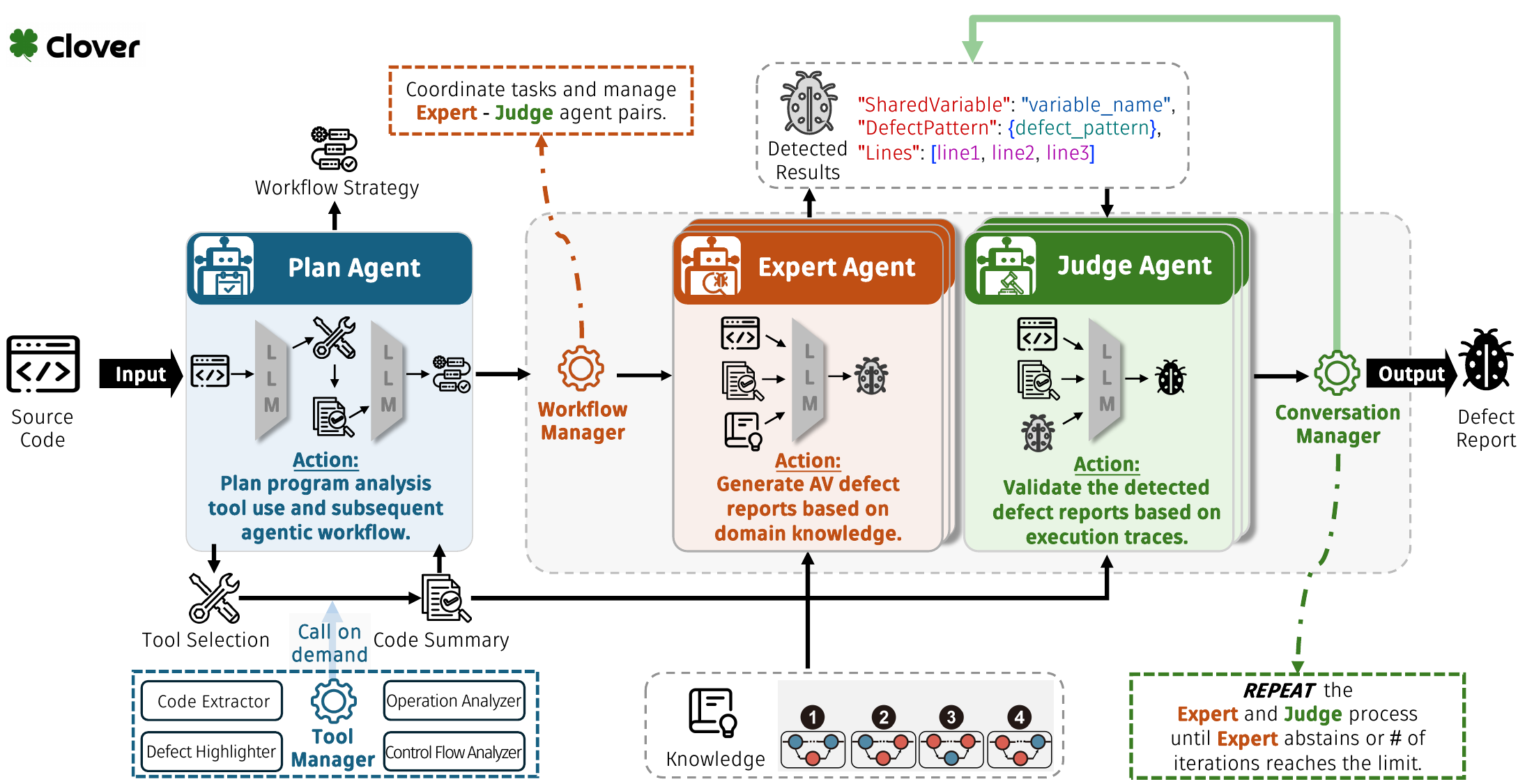}
    \vspace{-0.1in}
    \caption{Overview of \textbf{\tool}. 
     \footnotesize{\textnormal{(It is a multi-agent system with three specialized agents in a coordinated pipeline: (1) \emph{Plan Agent} performs code analysis via \emph{Tool Manager}, and orchestrates the \emph{Expert}-\emph{Judge} workflow, (2) \emph{Expert Agent} creates \emph{Expert}-\emph{Judge} agent pair via \emph{Workflow Manager} and detects atomicity violations using domain knowledge, (3) \emph{Judge Agent} validates defect reports through execution trace analysis. \emph{Expert} and \emph{Judge} agents iteratively collaborate via \emph{Conversation Manager} through multi-round dialogues for defect refinement and validation.)}} 
    }
    \label{fig:overview}
    \vspace{-5pt}
\end{figure*}

\tool is a multi-agent framework for detecting atomicity violations in real-world interrupt-driven programs, integrating static analysis tools with LLM agents. As illustrated in Figure~\ref{fig:overview}, \tool consists of three agents and four static analysis tools, coordinated with three rule-based managers.

Overall, \tool has three phases. In the first phase, the \emph{Plan Agent} dynamically activates the static analysis tools according to the given source code, with the help of \emph{Tool Manager}. It first invokes \emph{Operation Analyzer} to identify shared resources between ISRs and the related operations. For each shared resource, it applies \emph{Defect Highlighter} to capture the potential defect locations for further examination. If such location exists, it then analyze the complexity of the source code, invoking a \emph{Code Extractor} for code that exceed LLM context limit, and a \emph{Control Flow Analyzer} for code with complex function-call relationship. Next, the \emph{Tool Manager} then combine these information into a comprehensive code summary and decides the defect analysis strategy. Note that, \tool focus on only one shared variable at a time.

In the second phase, the \emph{Expert Agent} invoke the \emph{Workflow Manager} to generate a prompt that annotates the source code with code summary. Based on the the flagged shared variables and their access patterns, it also use the domain knowledge of corresponding atomicity violation patterns to equip the LLM, and generates candidate defect reports specifying the shared variable, defect pattern type, and suspicious code lines that may trigger violations.

In the third phase,  \emph{Judge Agent} analyzes the execution trace with interrupt-switch knowledge to validate the candidate defect report, judging whether each candidate violation is triggerable. Via \emph{Conversation Manager}, the \emph{Expert Agent} and \emph{Judge Agent} communicate in multiple rounds, iteratively refining the defect report until achieve consensus or reach the iteration limit. 

\subsection{Plan Agent}

\emph{Plan Agent} serves as the central orchestrator of \tool, coordinating static analysis tools and managing the overall analysis workflow. It analyzes source code, determines the optimal tool selection strategy, and provides comprehensive static analysis results for subsequent \emph{Expert-Judge} analysis. \emph{Plan Agent} operates through the \emph{Tool Manager} module, which registers and coordinates four static analysis tools designed for different tasks to obtain a comprehensive code summary. \emph{Plan Agent} integrates four specialized static analysis tools to achieve comprehensive code understanding:

\begin{algorithm}
\caption{\emph{Operation Analyzer}}
\label{alg:read-write-analyzer}
\scriptsize
\KwIn{C source file $S$}
\KwOut{Access matrix $A = \{(f, op, var, loc)\}$, $F_{\text{all}}$, $V_{\text{global}}$, $F_{\text{entry}}$, $G$}
\BlankLine

$IR \leftarrow \text{CompileToLLVMIR}(S)$\;\label{alg1:line1}
\If{$IR = \emptyset$}{\label{alg1:line2}
    \Return $\emptyset$\;\label{alg1:line3}
}

$F_{\text{all}} \leftarrow \text{ExtractFunctions}(IR)$\;\label{alg1:line5}
$V_{\text{global}} \leftarrow \text{ExtractGlobalVariables}(IR)$\;\label{alg1:line6}
$F_{\text{entry}} \leftarrow \text{FilterEntryFunctions}(F_{\text{all}})$\;\label{alg1:line7}
$G \leftarrow \text{BuildCallGraph}(F_{\text{all}})$\;\label{alg1:line8}

\ForEach{function $f \in F_{\text{all}}$}{\label{alg1:line10}
    \ForEach{instruction $I \in f$}{\label{alg1:line11}
        \If{\textnormal{IsGlobalAccess}$(I, V_{\text{global}})$}{\label{alg1:line12}
            $var \leftarrow \text{GetAccessedVariable}(I)$\;\label{alg1:line13}
            $op \leftarrow \text{GetOperationType}(I)$\;\label{alg1:line14}
            $loc \leftarrow \text{GetLocation}(I)$\;\label{alg1:line15}
            $A \leftarrow A \cup \{(f, op, var, loc)\}$\;\label{alg1:line16}
        }
    }
}
\Return $A$, $F_{\text{all}}$, $V_{\text{global}}$, $F_{\text{entry}}$, $G$\;\label{alg1:line18}
\end{algorithm}

\begin{algorithm}
\caption{\emph{Code Extractor}}
\label{alg:code-extractor}
\scriptsize
\KwIn{C source file $S$, $F_{\text{all}}$, $V_{\text{global}}$, $F_{\text{entry}}$, $G$}
\KwOut{Compressed code $S'$}
\BlankLine

$F_{\text{reach}} \leftarrow \emptyset$, $W \leftarrow F_{\text{entry}}$\;\label{alg2:line1}

\While{$W \neq \emptyset$}{\label{alg2:line2}
    $f_{\text{curr}} \leftarrow \text{SelectFunction}(W)$\;\label{alg2:line3}
    $W \leftarrow W \setminus \{f_{\text{curr}}\}$\;\label{alg2:line4}
    \If{$f_{\text{curr}} \notin F_{\text{reach}}$}{\label{alg2:line5}
        $F_{\text{reach}} \leftarrow F_{\text{reach}} \cup \{f_{\text{curr}}\}$\;\label{alg2:line6}
        $F_{\text{called}} \leftarrow \text{GetCallees}(G, f_{\text{curr}})$\;\label{alg2:line7}
        \ForEach{function $f_{\text{callee}} \in F_{\text{called}}$}{\label{alg2:line8}
            \If{$f_{\text{callee}} \notin F_{\text{reach}}$}{\label{alg2:line9}
                $W \leftarrow W \cup \{f_{\text{callee}}\}$\;\label{alg2:line10}
            }
        }
    }
}
$S' \leftarrow \text{GenerateExtractedCode}(S, F_{\text{reach}}, V_{\text{global}})$\;\label{alg2:line13}
\Return $S'$\;\label{alg2:line14}
\end{algorithm}

\paragraph{Operation Analyzer.} It compiles the C source file $S$ to LLVM IR and extracts functions $F_{\text{all}}$, global variables $V_{\text{global}}$, entry functions $F_{\text{entry}}$, and call graph $G$ (lines \ref{alg1:line1}-\ref{alg1:line8}). For each function $f \in F_{\text{all}}$ and instruction $I$, it identifies global accesses and records tuples $(f, op, var, loc)$ in the access matrix $A$ (lines \ref{alg1:line10}-\ref{alg1:line16}). For global variables of array type, it also determines their access patterns and valid index ranges with constant propagation and interval analysis~\cite{jaulin2000interval}. For pointers, it performs pointer-to-global mapping analysis to track indirect memory accesses\cite{10.1145/3374216} and resolve pointer aliasing relationships across function boundaries. For structure and unions, it analyzes field-level access patterns through \code{GetElementPtr} instruction decomposition~\cite{10.1145/3631882.3631885} to identify member-specific operations. Finally, it returns $A$, $F_{\text{all}}$, $V_{\text{global}}$, $F_{\text{entry}}$, and $G$ (line \ref{alg1:line18}) (Algorithm~\ref{alg:read-write-analyzer}).

\paragraph{Defect Highlighter.} 
It employs a sophisticated rule-based approach to identify potential atomicity violation locations through enhanced pattern matching. 
For a violation pattern $<a_1, a_2, a_3>$ on global variable $x$, it first identifies all functions containing both $a_1$ and $a_3$ operations on $x$. During this process, it conducts array index semantic equivalence analysis for dynamic access patterns, control-flow refinement to distinguish mandatory versus conditional accesses, and loop context detection with explicit basic block pattern recognition.
Next, it locates the code locations of $a_2$ operations on $x$ in higher-priority functions.
Finally, it validates whether each code location triples for $<a_1, a_2, a_3>$ are reachable in one execution path, through dominator tree analysis.

\paragraph{Code Extractor.} It takes source code $S$ and analysis results ($F_{\text{all}}$, $V_{\text{global}}$, $F_{\text{entry}}$, $G$) as input. For each unvisited function $f_{\text{curr}}$, it adds $f_{\text{curr}}$ to $F_{\text{reach}}$ (lines \ref{alg2:line5}-\ref{alg2:line6}) and discovers callees through call graph $G$ (line \ref{alg2:line7}), adding unvisited callees back to $W$ (lines \ref{alg2:line8}-\ref{alg2:line10}). Finally, it generates compressed code $S'$ by removing unreachable functions from $S$ while preserving all global variables $V_{\text{global}}$ and maintaining line number mapping information between original and compressed code for detection purposes (line \ref{alg2:line13}) (Algorithm~\ref{alg:code-extractor}).

\paragraph{Flow Analyzer.} It first conducts cross-file function-call analysis with LLVM, and filters out external library calls. It also identifies and tracks interrupt control points by locating interrupt disable/enable function calls (\eg, \code{disable_isr()} and \code{enable_isr()}). Finally, it reports function-call relationship and interrupt flow information.

\begin{figure*}
\centering
\includegraphics[width=0.8\linewidth]{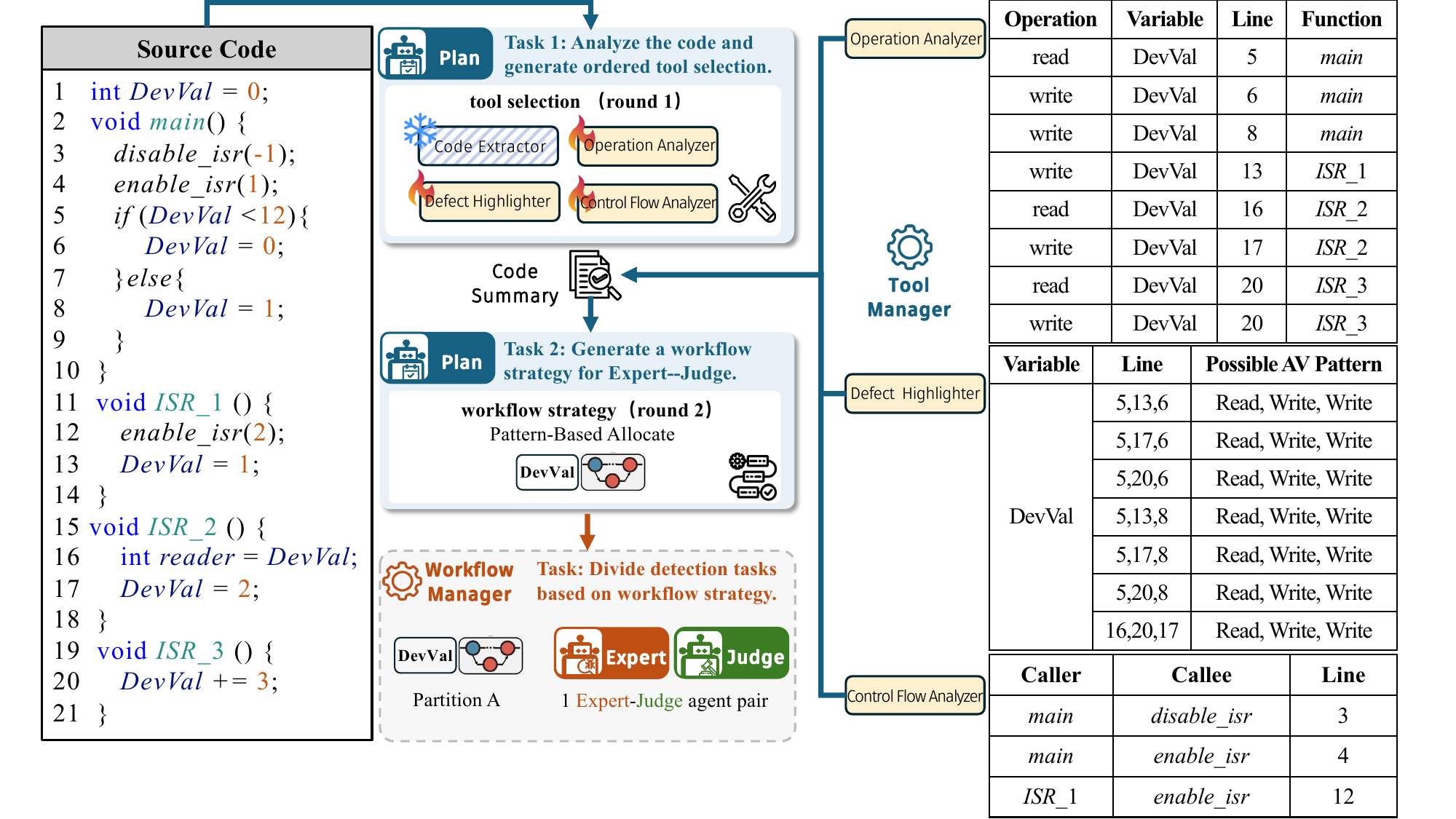}
\caption{The result of \emph{Plan Agent} and \emph{Workflow Manager} of the code in Figure~\ref{fig:motivation}.
}
\label{fig:example-3.1}
\vspace{-10pt}
\end{figure*}

\paragraph{Illustrative Example.} Consider the code in Figure~\ref{fig:motivation}. As illustrated in Figure~\ref{fig:example-3.1}, the \emph{Plan Agent} first invokes the tool manager to activate three static analysis tools. It identifies (1) 3 read operations and 5 write operations of \code{DevVal}; (2) 3 interrupt disable/enable function calls; and (3) 7 possible AV bugs of $<R, W, W>$ pattern. \emph{Code Extractor} is not invoked due to the small program size.

\subsection{Expert Agent}

The \emph{Expert Agent} is the core of \tool, which identifies atomicity violations through LLM-based analysis guided by the results of \emph{Plan Agent}. 
It contains two modules: workflow management and pattern-guided defect detection.

\subsubsection{Workflow Management.} \label{sec:worflow_management}
The \emph{Workflow Manager} transforms the static analysis results of \emph{Plan Agent} into concrete multi-agent deployment plans, partitioning the detection tasks (\ie, the highlighted locations) in a mutually exclusive and collectively exhaustive way. 

The \emph{Expert Agent} has two detection strategies. If there are less than 3 global shared variables and their potential violation patterns are roughly even-distributed, the pattern-based strategy is activated, which divides the detection tasks by patterns. Otherwise, the variable-based strategies activated, which divides the detection tasks by variables, where each group contains 1 high-frequency variable or at most 3 other variables. This dual-level grouping ensures comprehensive detection while preventing task overload for agents.

For each group of tasks, \emph{Workflow Manager} annotate each code statement that references the variables under detection, helping LLMs to concentrate. The annotation includes all the related static analysis results, in the form of code comments.
It also includes line number annotations to assist LLMs in describing code location.

\paragraph{Illustrative Example.} In Figure~\ref{fig:example-3.1}, the \emph{Workflow Manager} activates the pattern-based strategy, as only variable \code{DevVal} with pattern ($<R, W, W>$) is highlighted. It then deploys one specialized \emph{Expert-Judge} agent pair for this detection task.

\subsubsection{Pattern-Guided Defect Detection.} Next, the \emph{Expert Agent} performs the atomicity violation detection. To facilitate LLMs in understanding interrupt-driven programs, \tool constructs four knowledge modules, creating a pool of expert LLMs, each responsible for detecting a specific type of atomicity violation pattern. Each knowledge module provides a precise definition of the corresponding pattern (see Figure~\ref{fig:pattern}), accompanied by a representative code example.
During the analysis, \emph{Expert Agent} asynchronously invokes the relevant expert LLMs based on the workflow.

\begin{figure*}
\centering
\includegraphics[width=1.02\linewidth]{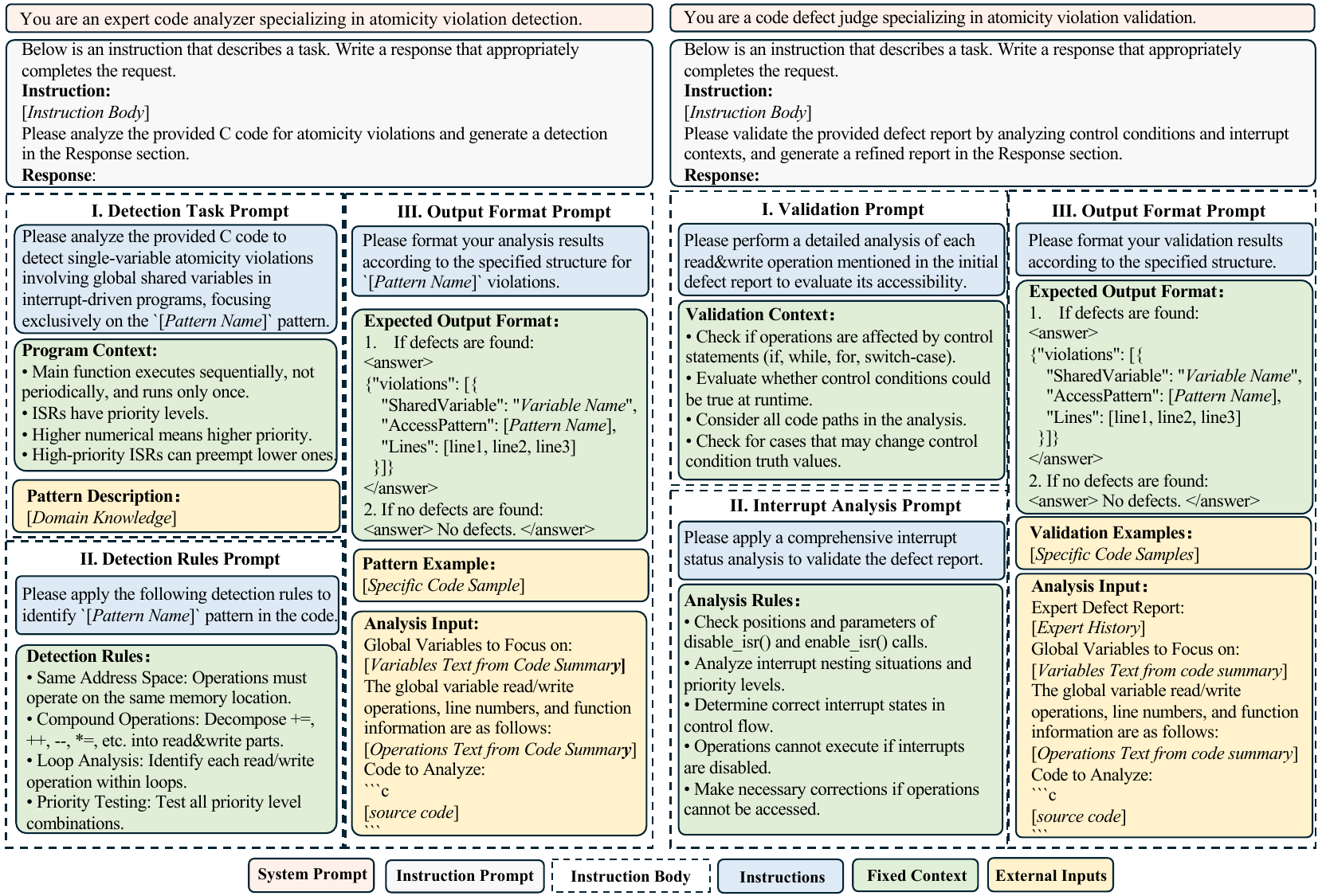}

(a) Expert Agent Prompt Template\qquad\qquad\qquad(b) Judge Agent Prompt Template

\vspace{-0.1in}
\caption{Prompt Template Design for Multi-Agent Atomicity Violation Detection.}
\label{fig:expert-judge-prompts}
\vspace{-10pt}
\end{figure*}

We carefully design the prompts for expert LLMs, incorporating implicit chain-of-thought reasoning~\cite{wei2022chain}, as shown in Figure~\ref{fig:expert-judge-prompts}. To enable the LLMs to handle variable references in diverse scenarios and their interactions between ISRs, \tool explicitly requires LLMs to analyze compound expressions (\eg, \code{++}, and \texttt{+=}), loops, and other control-flow structures, which native LLMs typically struggle to comprehend. Specifically, the prompt consists of three key components.

\begin{enumerate}[leftmargin=*]
\item \emph{Detection Task}: State the analysis objective and specify the atomicity violation pattern to focus on. Provide relevant program context and describe the pattern using domain knowledge module.
\item \emph{Detection Rules}: Apply step-by-step detection rules to identify the specified pattern in the code. Decompose compound operations, analyze read/write operations within loops, and test all priority level combinations as required.
\item \emph{Output Format}: Present the analysis results in the prescribed output format. Offer a one-shot example of the corresponding atomicity violation pattern, using case examples from RaceBench~\cite{racebench}\footnote{The case examples are not included in the \racebench dataset.}, and utilize the provided analysis input, which consists of a tool-generated code summary and annotated source code, to support accurate and thorough reasoning.

\end{enumerate}

\begin{figure}[]
\centering
\includegraphics[width=\linewidth]{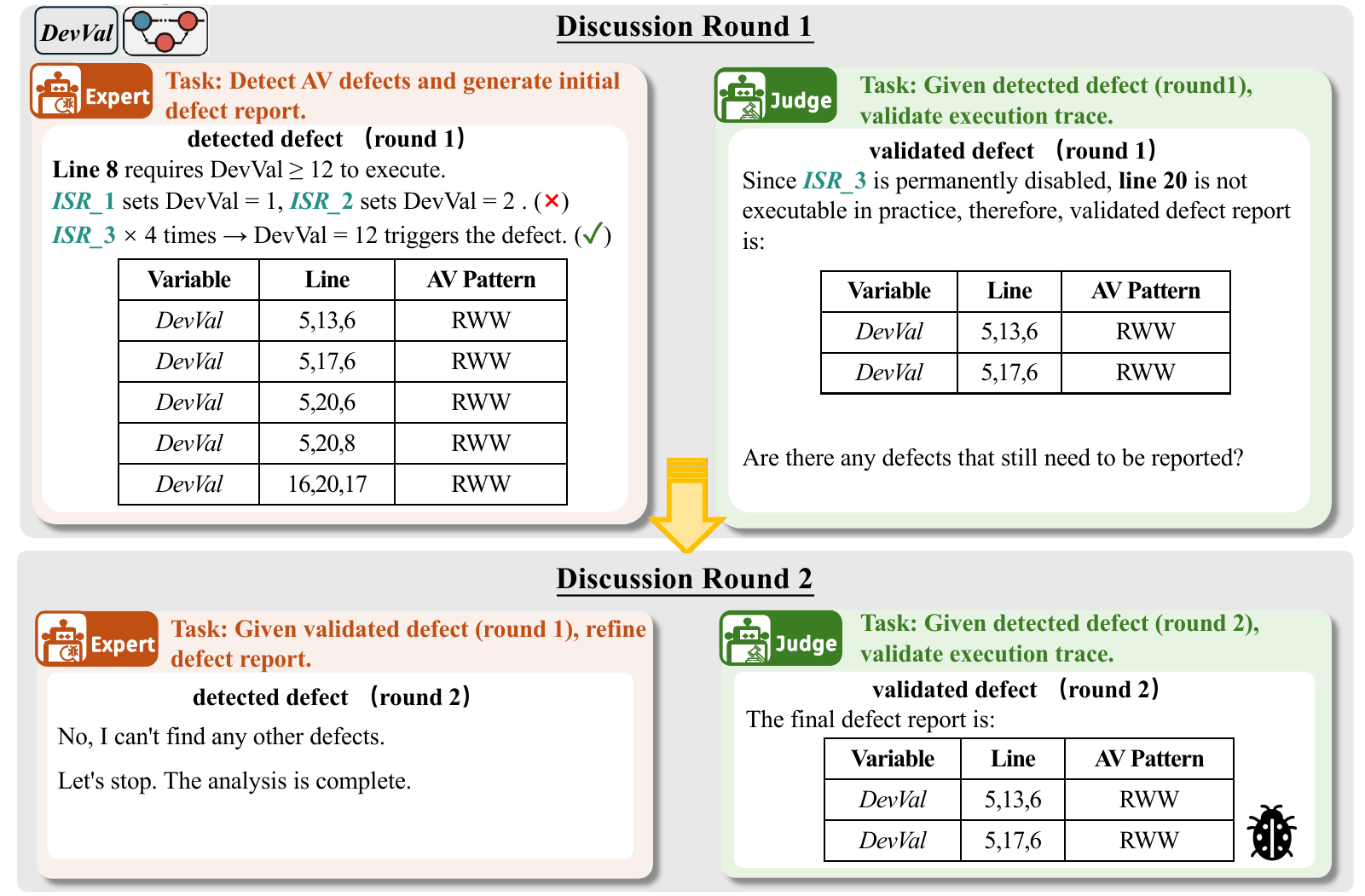}
\vspace{-0.1in}
\caption{The result of Expert Agent and Judge Agent of the code in Figure~\ref{fig:motivation}.
}
\vspace{-10pt}
\label{fig:example-3.2}
\end{figure}

\paragraph{Illustrative Example.} The upper-left part of Figure~\ref{fig:example-3.2} shows the result of \emph{Expert Agent}, for variable \texttt{DevVal} in Figure~\ref{fig:example-3.1}. \emph{Expert Agent} successfully identifies the path constraints for executing line 8, which requires \emph{ISR\_}1 and \emph{ISR\_}2 not preempting \code{main} function before the branch on line 5. Therefore, it removes two of the possible AV bugs highlighted by the \emph{Plan Agent}, and generates the initial defect report containing five potential $<R, W, W>$ atomicity violations.

\subsection{Judge Agent}

The non-deterministic execution order of interrupt-driven programs brings up a vast program state space. However, due to the interrupt priorities and satisfiability of control conditions, only a subspace is actually reachable during execution. To address this problem, \emph{Judge Agent} is proposed to analyze possible execution traces and validate the results provided by \emph{Expert Agent}.

The \emph{Judge Agent} focuses on the interrupt status and control conditions, and is equipped with the knowledge of control conditions and the interrupt-switching mechanism. Its prompt also follows implicit chain-of-thought reasoning, as shown in Figure~\ref{fig:expert-judge-prompts}. To enhance LLM's capability, \tool additionally provides descriptions and examples of interrupt status changes. Specifically, the prompt contains three key components:

\begin{enumerate}[leftmargin=*]
\item \emph{Validation}: Conduct a thorough analysis of each read and write operation identified in the initial defect report to assess its accessibility.
\item \emph{Interrupt Analysis}: Perform a comprehensive analysis of interrupt status to validate the defect report rigorously.
\item \emph{Output Format}: Provide the commented source code and the defect report from \emph{Expert Agent}.
\end{enumerate}

\begin{figure*}
    \centering
    \includegraphics[width=\linewidth]{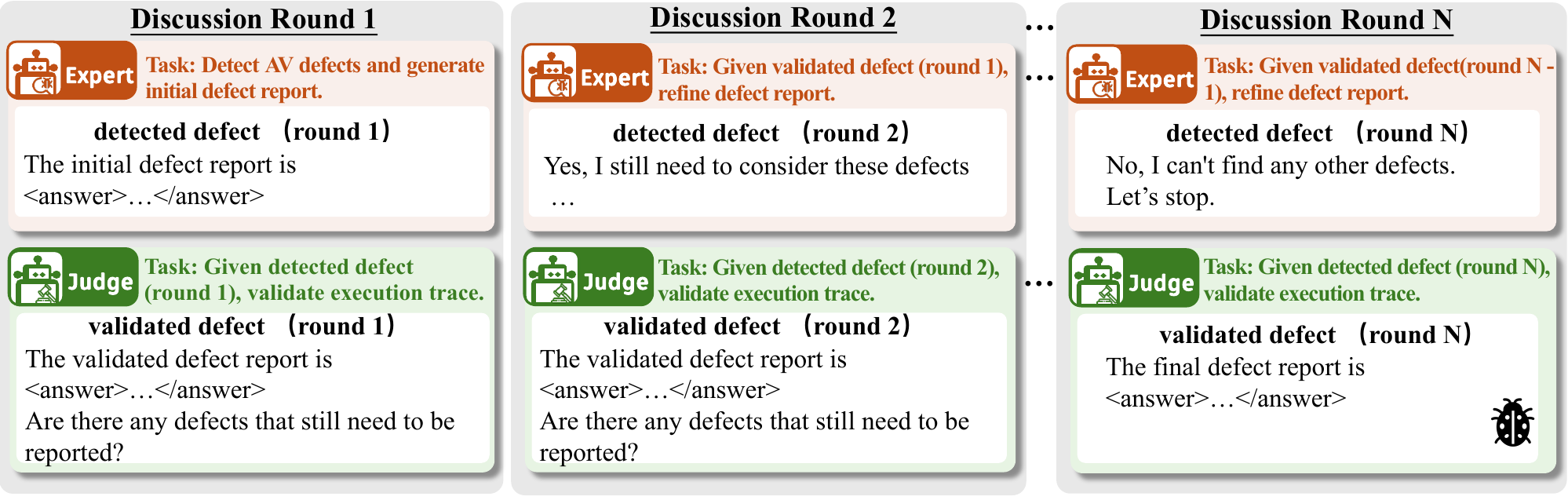}
    \vspace{-0.1in}
    \caption{An example of \tool's Expert-Judge multi-agent design.}
    \label{fig:agent-communication}
    \vspace{-5pt}
\end{figure*}

\paragraph{Multi-Agent Collaborative Process.} 
The \emph{Workflow Manager} creates an \emph{Expert-Judge} agent pair for each detection task partition, in an aim to prevent cross-contamination between  analysis sessions.
As shown in Figure~\ref{fig:agent-communication}, the \emph{Expert Agent} and \emph{Judge Agent} collaborate in an iterative process.
In each iteration, \emph{Expert Agent} first detects atomicity violations and generates the defect report (or refines the report based on feedback from the previous round). The \emph{Judge Agent} then validates whether these reported violations could be triggered in any reachable program state and provides validation feedback to the \emph{Expert Agent} to guide report refinement in the next round. 

\paragraph{Conversation Manager.} The multi-agent collaboration and the context management is handled by the \emph{Conversation Manager}, which contains three mechanisms: context curation, information integration, and iteration control.

The context curation maintains structured conversation histories with selective retention of the most recent expert-judge exchanges, preventing context overflow. Each history entry contains four key fields: \code{role}, \code{round}, \code{purpose}, and \code{content}. It also provides essential code context in all iterations, as introduce in Section~\ref{sec:worflow_management}.

The information integration employs structured prompt templates that dynamically combine static analysis outputs with conversational context from previous rounds. Standardizing terminology and operation descriptions across all agent interactions, it ensures comprehensive and consistently formatted information about the code under analysis, defect report changes, and the rationale behind previous judgments. 

The iteration control has three termination conditions: (1) the \emph{Expert Agent} report that no defect exists; (2) the \emph{Expert Agent} chooses to "abstain" from further analysis; and (3) the maximum iteration limit $N$ is reached. The \emph{Conversation Manager} tracks conversation state and terminates iteration when any of the condition is met. 

\paragraph{Illustrative Example.}  

Figure~\ref{fig:example-3.2} shows the two-round iteration of \emph{Expert Agent} and \emph{Judge Agent}. In the first round, \emph{Judge Agent} examines the execution trace and identifies that \emph{ISR_}3 is permanently disabled, making line 20 unreachable. Therefore, it removes three violations in the initial defect report. In the second round, the \emph{Expert Agent} conducts the detection with the knowledge of validation result. As no new atomicity violations detected, the \emph{Conversation Manager} terminates the multi-agent process and outputs the defect report.

\section{Implementation}

We have implemented \tool for C/C++ programs, the most popular programming languages for interrupt-driven software. The core algorithm of \tool could be easily generalized to other programming languages.

\tool uses Clang-10.0~\cite{Clang} and LLVM-10.0~\cite{LLVM} for static analysis, and all its agents utilize Claude-Sonnet-4~\cite{claude-4} as their underlying LLM. To ensure deterministic outputs, we set the model temperature to 0 and use greedy search decoding~\cite{wang2023boosting}. 
By default, the \emph{Expert-Judge} workflow executes a maximum of three iterations ($N=3$) to refine the detection results. 
In Section~\ref{sec:llm_impact}, we apply \tool to several advanced LLMs, including three closed-source LLMs, i.e., Claude-3.5-Sonnet~\cite{claude-3.5}, Gemini-2.5-Pro~\cite{comanici2025gemini}, and Qwen-Max~\cite{qwen25}, as well as three open-source LLMs, i.e., Kimi-K2~\cite{team2025kimi}, DeepSeek-V3~\cite{liu2024deepseek} and LongCat-Flash-Chat~\cite{longcatflashtechnicalreport}. 

All experiments have been conducted on a machine equipped with a 12-core Intel i5-12490F processor (3.00 GHz), 32 GB RAM, and 1TB GLOWAY TC3100 SSD (RAID 5). The system features a 1000 Mbps network connection with a twisted-pair port. The LLM inference has been performed remotely using the Aigcbest~\cite{aigcbest} cloud service.




\section{Evaluation}

Our evaluation aims to answer three key research questions:

\textbf{RQ1 (Effectiveness):} 
How effective is \tool in detecting atomicity violations in interrupt-driven programs, compared to state-of-the-art approaches?

\textbf{RQ2 (Scalability):} 
How well does \tool scale when applied to large-scale real-world programs?

\textbf{RQ3 (Ablation):} 
What are the individual contributions of \tool's core components to its overall performance?

\begin{table*}[!tb]
\centering
\caption{Statistics of Benchmarks}
\setlength{\tabcolsep}{12pt}
\resizebox{0.8\linewidth}{!}{%
\begin{threeparttable}[b]
\tiny
\begin{tabular}{llll|cccc}
\hline
Code Source & \multicolumn{3}{c|}{Benchmark} & \#Module & LoC & \#ISR & \#Vio \\ \hline
\multirow{4}{*}{Academy} & \multicolumn{3}{l|}{RaceBench 2.1~\cite{racebench}} & 31 & 1753 & 48 & 57 \\ \cline{2-8}
& \multicolumn{3}{l|}{SV-COMP~\cite{SVCOMP2022}   } & & & & \\
& \multicolumn{3}{l|}{\,\,\,\,\,\,\,\,ldv-races} & 5 & 211 & 5 & 32 \\
& \multicolumn{3}{l|}{\,\,\,\,\,\,\,\,goblint-reg.} & 77 & 2066 & 82 & 86 \\
& \multicolumn{3}{l|}{\,\,\,\,\,\,\,\,pthread} & 10 & 486 & 21 & 27 \\ \hline
\multirow{7}{*}{Industry} & \multicolumn{3}{l|}{RWIP~\cite{interruptdriven}} & & & & \\
& \multicolumn{3}{l|}{\,\,\,\,\,\,\,\,blink} & 3 & 578 & 9 & 34 \\
& \multicolumn{3}{l|}{\,\,\,\,\,\,\,\,brake} & 3 & 373 & 8 & 40 \\
& \multicolumn{3}{l|}{\,\,\,\,\,\,\,\,i2c\_pca\_isa} & 3 & 1975 & 9 & 61 \\
& \multicolumn{3}{l|}{\,\,\,\,\,\,\,\,i8xx\_tco} & 3 & 1063 & 9 & 5 \\
& \multicolumn{3}{l|}{\,\,\,\,\,\,\,\,logger} & 3 & 2671 & 8 & 6 \\
& \multicolumn{3}{l|}{\,\,\,\,\,\,\,\,wdt\_pci} & 3 & 3513 & 8 & 40 \\ \hline
\multicolumn{4}{c|}{Total} & 141 & 14689 & 207 & 388 \\ \hline
\end{tabular}
\begin{tablenotes}
    \item[*] LoC refers to lines of code. 
    \item[*] \#Vio refers to the number of all atomicity violations in the benchmark.
\end{tablenotes}
\end{threeparttable}
}
\label{tab:benchmark}
\end{table*}

\subsection{Experimental Design and Settings}

\subsubsection{Benchmark.} As summarized in Table~\ref{tab:benchmark}, we evaluate \tool on both academic and industry benchmarks.

\begin{itemize}
    \item \textbf{\racebench}~\cite{racebench}: A widely used benchmark for atomicity violations. It contains 31 interrupt-driven C programs with handcrafted atomicity violation defects. It is designed to simulate common atomicity violation patterns in real-world programs.
    \item \textbf{SV-COMP}~\cite{SVCOMP2022}: A software verification dataset that  covers various defect types. Following the evaluation settings of CPA4AV~\cite{yu2023detecting}, we adopt the \textit{ldv-races}, \textit{goblint regression}, and \textit{pthread} packages, transforming multi-threaded programs into interrupt-driven ones while preserving their code logic and structure.
    \item \textbf{RWIP}~\cite{interruptdriven}: A collection of 18 real-world interrupt-driven programs derived from embedded systems, including temperature logging devices, MSP430 micro-controllers, brake-by-wire systems, and i8xx chipsets.
\end{itemize}

\subsubsection{Baselines.} We compare \tool with three state-of-the-art approaches, each leveraging distinct techniques for atomicity violation detection.

\begin{itemize}
\item \textbf{intAtom}~\cite{li2022precise}: It conducts pattern-based static analysis and utilizes staged path pruning combined with constraint solving to efficiently reduce false positives.

\item \textbf{CPA4AV}~\cite{yu2023detecting}: It is a static analysis tool, utilizing control-flow graphs (CFGs) to identify interruption points where ISRs may interfere with tasks with higher priorities. 

\item \textbf{\purellm}~\cite{DRB-LLM}: It is a recent purely LLM-based approach designed for detecting data races in OpenMP programs. It uses the DRB-ML dataset from DataRaceBench~\cite{DataRaceBench} for in-context learning (ICL). We configure it with the same LLM settings as \tool, and strictly follow its original ICL methodology. To the best of our knowledge, we are the first to adopt LLM for atomicity violation detection. As \purellm represents the closest related work leveraging LLMs for similar defect detection tasks, we include it as one of the baselines.

\end{itemize}

\subsubsection{Metrics.}
An atomicity violation is considered successfully detected when both its defect pattern and the sequence of three operations (in the form of line numbers and operation types) are correctly reported. Based on this criterion, we categorize the results as follows:
\begin{itemize}[leftmargin=*]
    \item True Positive (TP): A correctly detected atomicity violation.
    \item False Positive (FP): An incorrectly reported atomicity violation.
    \item False Negative (FN): A ground-truth atomicity violation that is not detected by an approach.
\end{itemize}

These values are subsequently utilized to compute the three key evaluation metrics: precision($\frac{\mathrm{TP}}{\mathrm{TP}+\mathrm{FP}}$), recall ($\frac{\mathrm{TP}}{\mathrm{TP}+\mathrm{FN}}$), and F1-score ($2\,\frac{\text{precision}\cdot\text{recall}}{\text{precision}+\text{recall}}$).
Their average values are computed collectively across all defects, rather than averaging the metrics across benchmarks.

\begin{figure}[!tb]
    \centering
    \includegraphics[width=0.5\linewidth]{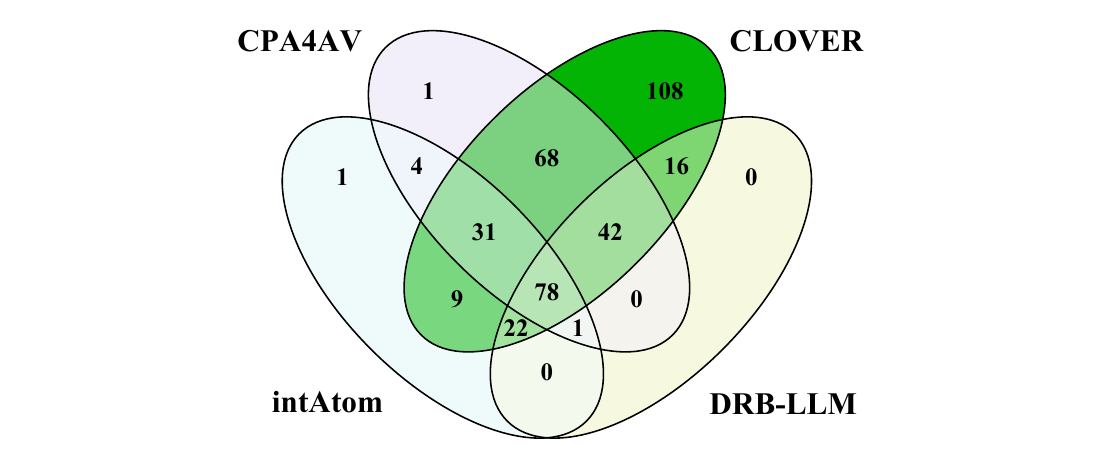}
    \caption{The correctly detected atomicity violations in across 3 benchmarks.}
    \label{fig:venn}
\end{figure}

\begin{table*}[]
\centering
\caption{Atomicity violation detection results across 3 benchmarks.}
\setlength{\tabcolsep}{5pt}
\resizebox{\linewidth}{!}{%
\begin{threeparttable}[b]
\footnotesize
\begin{tabular}{ll|cccc|cccc|cccc|cccc}
\hline
\multicolumn{2}{c|}{\multirow{2}{*}{Benchmark}} & \multicolumn{4}{c|}{\tool}                 & \multicolumn{4}{c|}{\purellm} & \multicolumn{4}{c|}{intAtom}                     & \multicolumn{4}{c}{CPA4AV}                                    \\ \cline{3-18} 
\multicolumn{2}{c|}{}                              & Pre        & Rec        & F1        & Time(s) & Pre         & Rec & F1 & Time(s) & Pre        & Rec & F1 & Time(s)      & Pre        & Rec       & F1        & Time(s)      \\ \hline
\multicolumn{2}{l|}{\racebench}     & 91.7           & \textbf{96.5}  & \textbf{94.0} & 45.0    & 38.9            & 77.2    & 51.8   & 20.0    & 89.5           & 89.5    & 89.5   & 0.1          & \textbf{100.0} & 45.6          & 62.7          & \textbf{0.1} \\ \hline
\multicolumn{2}{l|}{SV-COMP}                       & 96.0          & \textbf{97.9}  & \textbf{96.9} & 33.9    & 52.9            & 67.5    & 53.9   & 10.7    & \textbf{100.0} & 65.5    & 79.2   & \textbf{0.1} & 83.8           & 91.0          & 87.2          & 0.1          \\
\multicolumn{2}{l|}{\,\,\,\,\,\,\,\,ldv-races}     & \textbf{100.0} & \textbf{100.0}  & \textbf{100.0} & 75.7   & 82.4  & 87.5    & 84.8   & 16.0    & \textbf{100.0} & 21.9    & 35.9   & \textbf{0.2} & 61.5           & \textbf{100.0}        & 76.2          & 0.3          \\
\multicolumn{2}{l|}{\,\,\,\,\,\,\,\,goblint-reg.}  & 95.5           & \textbf{97.7}  & \textbf{96.6} & 27.7    & 27.0            & 69.8    & 39.0   & 12.4    & \textbf{100.0} & 81.4    & 89.7   & \textbf{0.1} & 88.1           & 86.0          & 87.1          & 0.1          \\
\multicolumn{2}{l|}{\,\,\,\,\,\,\,\,pthread}       & 92.9 & \textbf{96.3}           & 94.5          & 61.4   & 45.0            & 64.3    & 52.9   & 13.8    & \textbf{100.0} & 66.7    & 80.0   & \textbf{0.1} & 96.3 & \textbf{96.3} & \textbf{96.3} & 0.5          \\ \hline
\multicolumn{2}{l|}{RWIP}                          & \textbf{85.0}  & \textbf{95.2}  & \textbf{90.3} & 116.7   & 46.9            & 13.4    & 19.3   & 16.3    & 0              & 0       & 0      & \textbf{0.3} & 75.9           & 35.5          & 48.4          & 157.7        \\
\multicolumn{2}{l|}{\,\,\,\,\,\,\,\,blink}         & \textbf{97.1}  & \textbf{97.1}  & \textbf{97.1} & 106.9   & 60.0            & 17.6    & 27.3   & 13.9    & 0              & 0       & 0      & \textbf{0.2} & 88.5           & 67.6          & 76.7          & 5.1          \\
\multicolumn{2}{l|}{\,\,\,\,\,\,\,\,brake}         & \textbf{97.6} & \textbf{97.6}  & \textbf{97.6} & 140.4   & 21.4            & 7.5     & 11.1   & 19.4    & 0              & 0       & 0      & \textbf{0.3} & 85.7           & 15.0          & 25.5          & 600.4        \\
\multicolumn{2}{l|}{\,\,\,\,\,\,\,\,i2c\_pca    \_isa} & 90.2           & \textbf{90.2}  & \textbf{90.2} & 120.2   & 92.3  & 19.7    & 32.4   & 14.7    & 0              & 0       & 0      & \textbf{0.1} & \textbf{100.0} & 19.7          & 32.9          & 315.9        \\
\multicolumn{2}{l|}{\,\,\,\,\,\,\,\,i8xx\_tco}     & 62.5 & \textbf{100.0}  & \textbf{76.9} & 49.9   & 16.7            & 20.0    & 18.2   & 18.0    & 0              & 0       & 0      & \textbf{0.7} & \textbf{100.0} & 40.0          & 57.1          & 2.0          \\
\multicolumn{2}{l|}{\,\,\,\,\,\,\,\,logger}        & \textbf{85.7}  & \textbf{100.0} & \textbf{92.3} & 102.4   & 17.6            & 50.0    & 26.1   & 14.6    & 0              & 0       & 0      & \textbf{0.1} & 19.0           & 66.7          & 29.6          & 6.8          \\
\multicolumn{2}{l|}{\,\,\,\,\,\,\,\,wdt\_pci}      & 66.1           & \textbf{97.5}  & \textbf{78.8} & 180.2   & 0               & 0       & 0      & 17.4    & 0              & 0       & 0      & \textbf{0.4} & \textbf{100.0}         & 47.5          & 64.4          & 16.2         \\ \hline
\multicolumn{2}{c|}{Total/Average}                 & \textbf{91.0}  & \textbf{96.4}  & \textbf{93.6} & 37.0    & 44.3            & 45.4    & 44.8   & 14.8    & 50.5           & 37.6    & 43.1   & \textbf{0.1} & 90.2           & 57.7          & 70.4          & 20.2         \\ \hline
\end{tabular}%
    \begin{tablenotes}
      \item[*]"Pre(\%)" refers to Precision, "Rec(\%)" refers to Recall, and "F1(\%)" refers to the F1-score. \textbf{Bold numbers} indicate the best results.
      
    \end{tablenotes}
\end{threeparttable}
}
\label{tab:overall-result}
\end{table*} 

\begin{table*}[]
\caption{Detailed detection results on \racebench.}
\label{racebenchresult}
\footnotesize
\resizebox{1\linewidth}{!}{
\begin{threeparttable}[b]
\begin{tabular}{ccccc|cccc|cccc|cccc|cccc}
\hline
                       &                         &                         &                        &                         & \multicolumn{4}{c|}{\tool}                                                                       & \multicolumn{4}{c|}{\purellm} & \multicolumn{4}{c|}{intAtom} & \multicolumn{4}{c}{CPA4AV}   \\ \cline{6-21} 
\multirow{-2}{*}{\#ID} & \multirow{-2}{*}{\#Loc} & \multirow{-2}{*}{\#ISR} & \multirow{-2}{*}{\#SV} & \multirow{-2}{*}{\#Vio} & \#TP                      & \#FP                     & \#FN                     & Time(s)                       & \#TP     & \#FP     & \#FN     & Time(s)     & \#TP & \#FP & \#FN & Time(s) & \#TP & \#FP & \#FN & Time(s) \\ \hline
Ex1                    & 65                      & 2                       & 3                      & 2                       & { 2}  & { 0} & { 0} & { 44.7}   & 2        & 1        & 0        & 14.2        & 1    & 0    & 1    & 0.1     &      &      &      &         \\
Ex2                    & 46                      & 2                       & 1                      & 1                       & { 1}  & { 0} & { 0} & { 28.7}   & 1        & 2        & 0        & 13.6        & 1    & 0    & 0    & 0.1     &      &      &      &         \\
Ex3                    & 74                      & 2                       & 3                      & 1                       & { 1}  & { 0} & { 0} & { 35.7}   & 1        & 4        & 0        & 14.6        & 1    & 2    & 0    & 0.3     & 1    & 0    & 0    & 1.5     \\
Ex4                    & 69                      & 2                       & 4                      & 1                       & { 1}  & { 0} & { 0} & { 28.1}   & 1        & 5        & 0        & 21.2        & 1    & 0    & 0    & 0.3     & 1    & 0    & 0    & 0.1     \\
Ex5                    & 47                      & 1                       & 1                      & 1                       & { 1}  & { 0} & { 0} & { 28.3}   & 1        & 2        & 0        & 11.9        & 1    & 1    & 0    & 0.1     & 1    & 0    & 0    & 0.2     \\
Ex6                    & 54                      & 1                       & 2                      & 1                       & { 1}  & { 0} & { 0} & { 27.5}   & 1        & 3        & 0        & 12.6        & 1    & 2    & 0    & 0.3     & 1    & 0    & 0    & 0.3     \\
Ex7                    & 51                      & 1                       & 2                      & 2                       & { 2}  & { 0} & { 0} & { 81.0}   & 1        & 2        & 1        & 13.4        & 2    & 0    & 0    & 0.1     &      &      &      &         \\
Ex8                    & 53                      & 1                       & 1                      & 1                       & { 1}  & { 0} & { 0} & { 40.8}   & 1        & 3        & 0        & 14.7        & 1    & 0    & 0    & 0.1     &      &      &      &         \\
Ex9                    & 48                      & 1                       & 2                      & 3                       & { 3}  & { 0} & { 0} & { 72.3}   & 0        & 2        & 3        & 15.1        & 3    & 0    & 0    & 0.1     &      &      &      &         \\
Ex10                   & 54                      & 1                       & 2                      & 1                       & { 1}  & { 0} & { 0} & { 27.5}   & 1        & 1        & 0        & 12.6        & 1    & 0    & 0    & 0.0     &      &      &      &         \\
Ex11                   & 44                      & 1                       & 1                      & 1                       & { 1}  & { 0} & { 0} & { 23.8}   & 1        & 3        & 0        & 14.5        & 1    & 0    & 0    & 0.0     &      &      &      &         \\
Ex12                   & 35                      & 1                       & 1                      & 1                       & { 1}  & { 0} & { 0} & { 20.2}   & 1        & 1        & 0        & 10.8        & 1    & 0    & 0    & 0.0     &      &      &      &         \\
Ex13                   & 67                      & 3                       & 4                      & 1                       & { 1}  & { 0} & { 0} & { 47.4}   & 1        & 3        & 0        & 14.3        & 1    & 0    & 0    & 0.1     &      &      &      &         \\
Ex14                   & 60                      & 3                       & 4                      & 1                       & { 1}  & { 0} & { 0} & { 47.3}   & 1        & 5        & 0        & 15.8        & 1    & 0    & 0    & 0.1     &      &      &      &         \\
Ex15                   & 41                      & 1                       & 2                      & 1                       & { 1}  & { 0} & { 0} & { 23.6}   & 1        & 2        & 0        & 12.3        & 1    & 0    & 0    & 0.0     & 1    & 0    & 0    & 0.1     \\
Ex16                   & 34                      & 1                       & 1                      & 3                       & { 3}  & { 0} & { 0} & { 48.0}   & 3        & 1        & 0        & 12.3        & 3    & 0    & 0    & 0.0     & 3    & 0    & 0    & 0.0     \\
Ex17                   & 42                      & 1                       & 2                      & 5                       & { 3}  & { 2} & { 2} & { 123.9}  & 2        & 1        & 3        & 13.2        & 5    & 0    & 0    & 0.1     & 4    & 0    & 1    & 0.3     \\
Ex18                   & 65                      & 2                       & 2                      & 3                       & { 3}  & { 1} & { 0} & { 48.2}   & 3        & 2        & 0        & 13.9        & 3    & 0    & 0    & 0.0     &      &      &      &         \\
Ex19                   & 66                      & 1                       & 6                      & 2                       & { 2}  & { 1} & { 0} & { 38.2}   & 2        & 2        & 0        & 13.1        & 0    & 0    & 2    & 0.4     &      &      &      &         \\
Ex20                   & 55                      & 2                       & 3                      & 3                       & { 3}  & { 0} & { 0} & { 27.9}   & 2        & 3        & 1        & 15.2        & 1    & 0    & 2    & 0.1     &      &      &      &         \\
Ex21                   & 81                      & 1                       & 1                      & 3                       & { 3}  & { 0} & { 0} & { 107.0}  & 3        & 0        & 0        & 13.8        & 3    & 1    & 0    & 0.3     & 3    & 0    & 0    & 0.1     \\
Ex22                   & 67                      & 1                       & 1                      & 4                       & { 4}  & { 0} & { 0} & { 83.4}   & 3        & 1        & 1        & 15.3        & 4    & 0    & 0    & 0.1     & 4    & 0    & 0    & 0.2     \\
Ex23                   & 40                      & 1                       & 1                      & 2                       & { 2}  & { 0} & { 0} & { 72.0}   & 1        & 0        & 1        & 12.2        & 2    & 0    & 0    & 0.0     & 2    & 0    & 0    & 0.0     \\
Ex24                   & 65                      & 1                       & 1                      & 1                       & { 1}  & { 0} & { 0} & { 34.8}   & 1        & 1        & 0        & 12.3        & 1    & 0    & 0    & 0.1     &      &      &      &         \\
Ex25                   & 39                      & 1                       & 1                      & 1                       & { 1}  & { 0} & { 0} & { 26.8}   & 1        & 0        & 0        & 13.0        & 1    & 0    & 0    & 0.0     &      &      &      &         \\
Ex26                   & 44                      & 2                       & 1                      & 2                       & { 2}  & { 0} & { 0} & { 33.9}   & 2        & 3        & 0        & 14.1        & 2    & 0    & 0    & 0.0     & 1    & 0    & 1    & 0.0     \\
Ex27                   & 49                      & 3                       & 1                      & 3                       & { 3}  & { 0} & { 0} & { 34.4}   & 2        & 2        & 1        & 13.4        & 2    & 0    & 1    & 0.1     & 2    & 0    & 1    & 0.0     \\
Ex28                   & 54                      & 3                       & 2                      & 1                       & { 1}  & { 0} & { 0} & { 32.8}   & 1        & 6        & 0        & 18.3        & 1    & 0    & 0    & 0.1     & 1    & 0    & 0    & 0.0     \\
Ex29                   & 95                      & 1                       & 4                      & 1                       & { 1}  & { 0} & { 0} & { 28.7}   & 1        & 2        & 0        & 15.3        & 1    & 0    & 0    & 0.4     &      &      &      &         \\
Ex30                   & 57                      & 3                       & 2                      & 1                       & { 1}  & { 0} & { 0} & { 34.4}   & 1        & 5        & 0        & 197.1       & 1    & 0    & 0    & 0.1     & 1    & 0    & 0    & 0.1     \\
Ex31                   & 92                      & 1                       & 1                      & 3                       & { 3}  & { 1} & { 0} & { 43.1}   & 1        & 1        & 2        & 14.8        & 3    & 0    & 0    & 0.2     &      &      &      &         \\ \hline
Total                  & 1753                    & 48                      & 63                     & 57                      & { 55} & { 5} & { 2} & { 1394.0} & 44       & 69       & 13       & 619.0       & 51   & 6    & 6    & 3.6     & 26   & 0    & 3    & 3.1     \\ \hline
\end{tabular}
\begin{tablenotes}
     \item[*] The blank cell refers to the failed execution of a tool.
    \end{tablenotes}
\end{threeparttable}
    }
\end{table*}

\subsection{Answer to RQ1: Effectiveness}
\label{sec:eval_rq1}

Table~\ref{tab:overall-result} presents a summary of \tool's performance compared to baselines across the three benchmarks with ground-truth. Overall, \tool detects 96.4\% of atomicity violations while maintaining a low false positive rate of 9.0\%. It achieves an average F1-score of 93.6\%, significantly outperforming the baselines, which range from 43.1\% to 70.4\%. 

In comparison, CPA4AV, the only baseline with comparable precision as \tool, only detects 58.6\% of the atomicity violations detected by \tool. This limitation arises because CPA4AV struggles with complex data structure, which is commonly found in real-world programs (\eg, RWIP). In addition, it suffers program state space explosion problem, preventing it from thoroughly analyzing the program's execution states. Similarly, intAtom cannot interpret the function pointers and other complex data structures, leading to empty defect reports on all programs in RWIP. Meanwhile, \purellm lacks domain knowledge of interrupt-driven programs and fails to capture long-range program dependencies, leading to the worst performance. 

\paragraph{Atomicity violation detection.} Table~\ref{tab:benchmark} and Figure~\ref{fig:venn} summarizes the correctly detected atomicity violations of each approach. Among them, 124 violations are not detected by any of the static analysis approaches (\ie, intAtom and CPA4AV) and 222 violations are missed by the LLM-based approach (\ie, \purellm). In contrast, \tool detects the majority of them (374 out of 388), highlighting the advantage of its hybrid methodology. Furthermore, \tool detects 108 defects that none of the baselines is able to discover.

\begin{figure}
    \centering
    \begin{subfigure}[b]{0.48\textwidth}
        \centering
        \includegraphics[width=\textwidth]{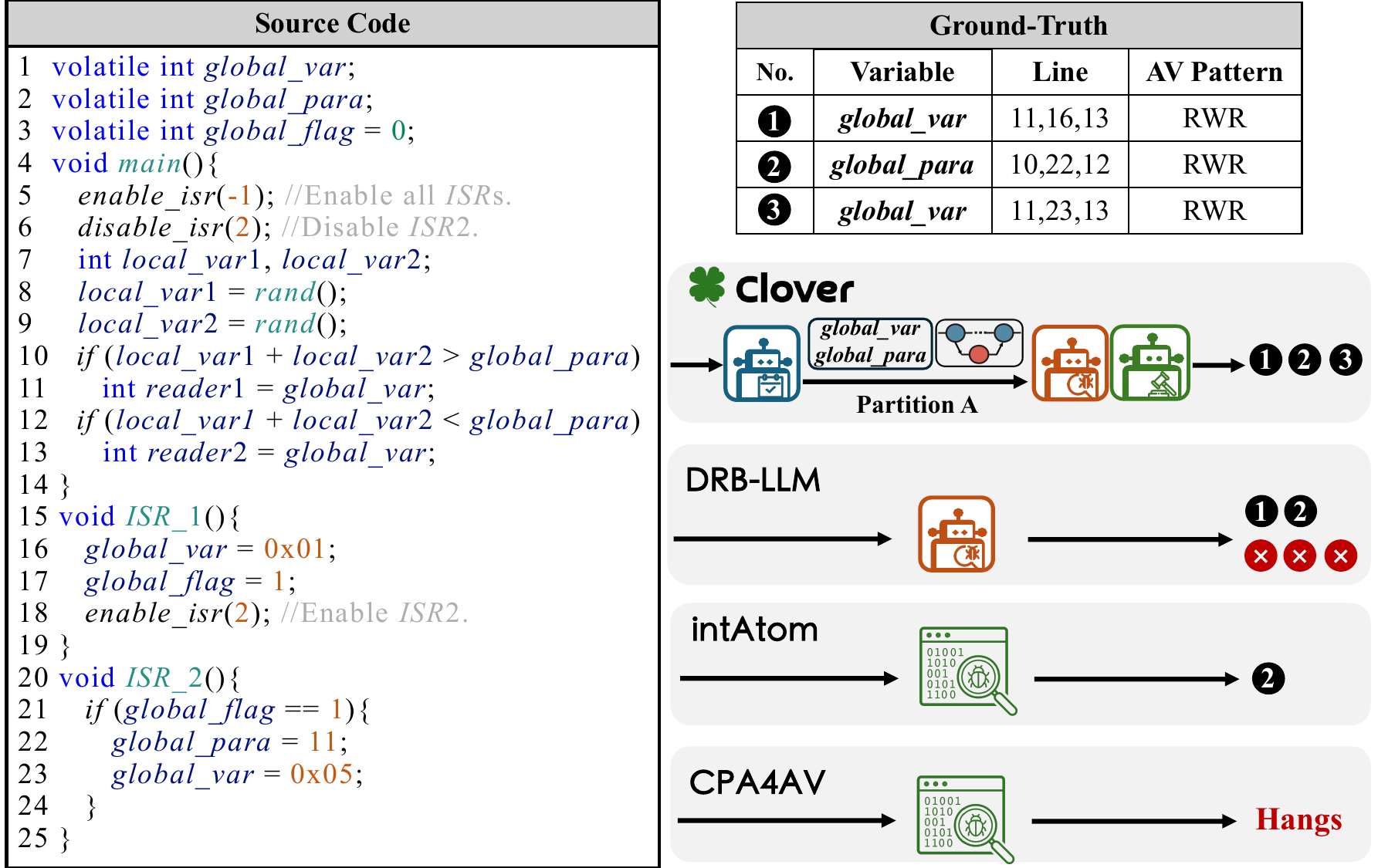}
        \caption{Result of Ex20 from \racebench.}
        \label{fig:RQ1-example}
    \end{subfigure}
    \hfill
    \begin{subfigure}[b]{0.48\textwidth}
        \centering
        \includegraphics[width=\textwidth]{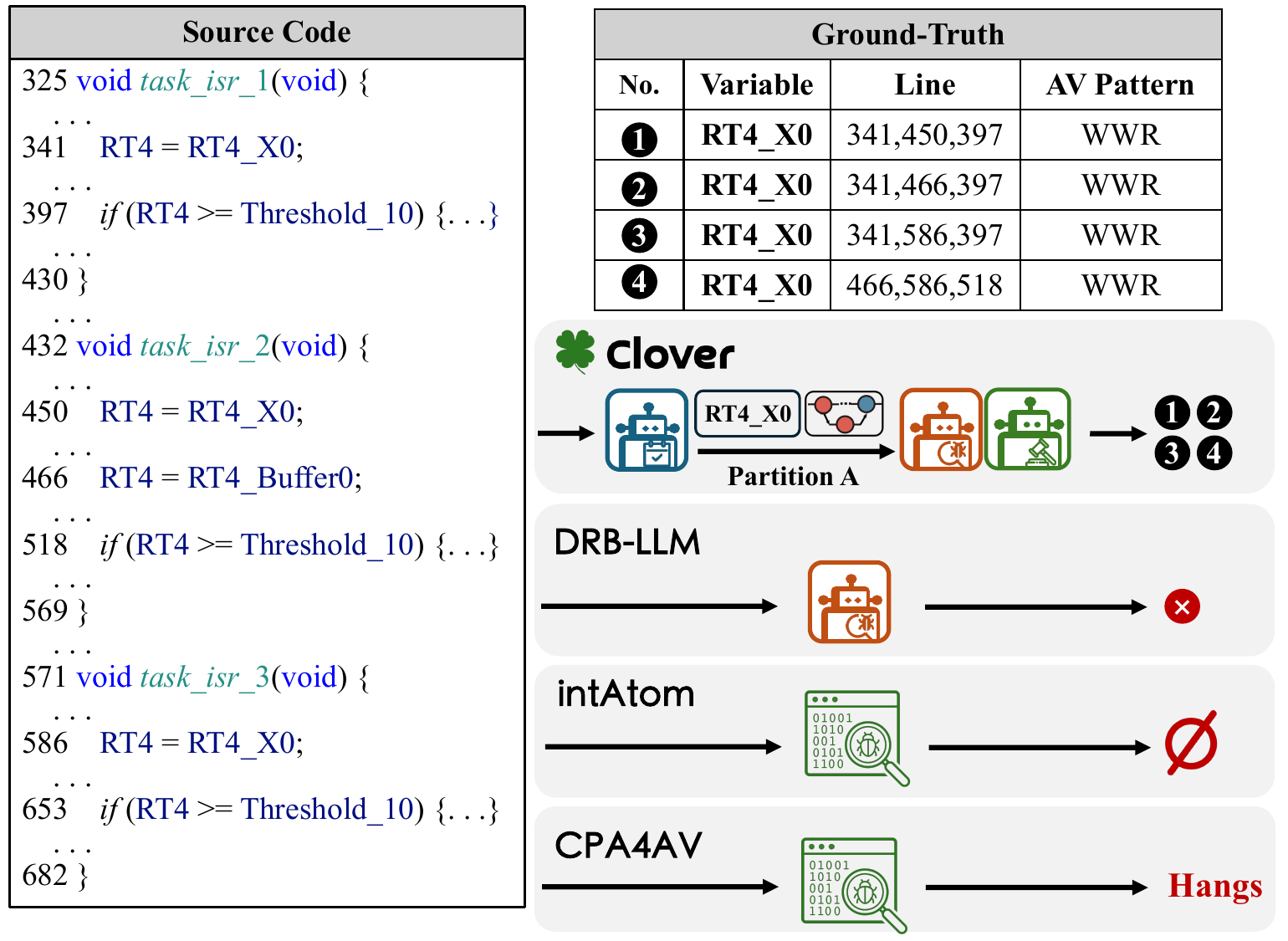}
        \caption{Result of a snippet from \code{brake} package of RWIP.}
        \label{fig:RQ3-example-1}
    \end{subfigure}
    \caption{Example results showing \tool's detection capabilities on different benchmark datasets.}
    \label{fig:examples}
\end{figure} 

\paragraph{Time consumption.} Due to network latency and the computation cost of LLM inference in multi-agent processes, \tool exhibits the highest time consumption on small-scale programs. However, as program scale and complexity increases, the execution time of static analysis approaches grows significantly. This trend is particularly evident in the \code{brake} package (the largest one in RWIP), where the execution time of CPA4AV surpasses that of \tool. Note that, \purellm is faster than \tool because it performs a simpler task with a more straightforward design, as well as having less details in the defect reports. However, its lower computational cost comes at the expense of accuracy, achieving the lowest F1-score despite being evaluated under a more lenient criterion.

\paragraph{Case Study.} Table~\ref{racebenchresult} details the performance of each approach on \racebench. 
\tool achieves the highest recall (96.4\%) and F1-score (93.6\%). IntAtom performs competitively in this benchmark as it is specifically designed for the \racebench and SV-COMP benchmark.
In contrast, CPA4AV, which supports only a subset of code syntax, fails to execute on 17 programs. Meanwhile, \purellm struggles to understanding atomicity violations, producing false positives in nearly all test cases.

Figure~\ref{fig:RQ1-example} illustrates an example containing three atomicity violation defects. \tool first identifies two critical resources, \code{global_var} and \code{global\_para}, and extracts corresponding code snippets. Through the multi-agent process, \tool accurately interprets the control-flow structure and detects all three defects. 
In contrast, intAtom detects only one defect, as it fails to identify that the conditions in line 10 and line 12 could be both fulfilled in the same run due to interruptions.

Similarly, \purellm misinterprets the control flow, resulting in one missed defect and three false positives. 
Additionally, CPA4AV encounters state space explosion problems and fails to complete execution. Such state space explosion is caused by the countless possible interleavings of conditional branches and the preemption points of ISRs.

\begin{tcolorbox}[size=title,colframe=white,width=1\linewidth,colback=gray!20]
\textbf{Summary}. \tool is highly effective in detecting atomicity violations. It achieves a precision/recall of 91.0\%/96.4\%, outperforming existing approaches by 33.0-117.2\% on F1-score.
\end{tcolorbox}

\subsection{Answer to RQ2: Scalability}

To assess the scalability of \tool, we evaluate \tool on two real-world interrupt-driven benchmarks: RWIP and Aerospace.

\paragraph{RWIP benchmark.} As summarized in Table~\ref{tab:overall-result}, \tool successfully detects 95.2\% of atomicity violations while having a low false positive rate of 13\%, demonstrating its effectiveness in analyzing large-scale real-world programs. In contrast, \purellm encounters context length limitations, preventing it from processing several larger programs. IntAtom fails to comprehend complex control flow structures and extensive function call graphs, resulting in empty defect reports for all programs. While CPA4AV achieves a precision comparable to \tool, it only identifies 42.1\% defects,  as it struggles with complex data structures and code logic.

Figure~\ref{fig:RQ3-example-1} illustrates a representative case containing four $<R,W,R>$ atomicity violations related to the variable \code{RT4_X0}. \tool identifies critical resources and extracts the core code, allowing it to accurately and efficiently capture the interactions between the three ISRs operating on \code{RT4_X0}. Consequently, it  detects all four atomicity violations without  any false positives. In contrast, \purellm fails to distinguish between the global variables with similar identifiers (\eg, \code{RT4\_X0} and \code{RT4\_Buffer0}). It fails to find any defects, and even has one incorrect report. Due to the complex control flow structure and huge function call graph, intAtom fails to comprehend the program and does not report any defect. CPA4AV terminates abnormally due to the state space explosion problem.

\begin{tcolorbox}[size=title,colframe=white,width=1\linewidth,colback=gray!20]
\textbf{Summary}: \tool demonstrates strong scalability, effectively detecting atomicity violations in large-scale real-world applications. In contrast, the baselines either fail to execute or ignore many defects.
\end{tcolorbox}

\subsection{Answer to RQ3: Ablation}

\begin{table*}[]
\centering
\caption{Ablation study results of \tool. }
\setlength{\tabcolsep}{5pt}
\resizebox{\linewidth}{!}{
\begin{threeparttable}[b]
\begin{tabular}{ll|cccc|cccc|cccc|cccc}
\hline
\multicolumn{2}{c|}{\multirow{2}{*}{Benchmark}} & \multicolumn{4}{c|}{\tool}                 & \multicolumn{4}{c|}{(A) \tool w/o JA}                & \multicolumn{4}{c|}{(B) \tool w/o PA} & \multicolumn{4}{c}{(C) \tool w/o PA \& JA} \\ \cline{3-18} 
\multicolumn{2}{c|}{}                              & Pre        & Rec        & F1        & Time(s) & Pre        & Rec       & F1        & Time(s)        & Pre   & Rec   & F1  & Time(s)        & Pre    & Rec    & F1    & Time(s)         \\ \hline
\multicolumn{2}{c|}{RaceBench 2.1}                 & \textbf{91.7}  & \textbf{96.5}  & \textbf{94.0} & 45.0    & 70.1           & 94.7          & 80.6          & \textbf{18.8}  & 71.7      & 66.7      & 69.1    & 37.4          & 44.4       & 64.3       & 52.2      & 26.0           \\ \hline
\multicolumn{2}{l|}{SV-COMP}                       &  \textbf{96.0}           & \textbf{97.9}  & \textbf{96.9} & 33.9    & 77.1  & 95.2          & 85.2          & \textbf{25.0}  & 53.5      & 62.8      & 57.7    & 31.9          & 44.5       & 60.7       & 51.4      & \textbf{19.3}            \\
\multicolumn{2}{l|}{\,\,\,\,\,\,\,\,ldv-races}     & \textbf{100.0}           & \textbf{100.0} & \textbf{100.0} & 75.7    & 60.8 & 96.9 & 74.7 & 55.7           & 54.1      & 62.5      & 58.0    & 30.3  & 39.6       & 59.4       & 47.5      & \textbf{23.7}          \\
\multicolumn{2}{l|}{\,\,\,\,\,\,\,\,goblint-reg.}  & \textbf{95.5}           & \textbf{97.7}  & \textbf{96.6} & 27.7    & 82.0           & 95.3          & 88.2          & \textbf{22.1}  & 50.5      & 64.0      & 56.4    & 30.2         & 49.1       & 62.8       & 55.1      & 30.2            \\
\multicolumn{2}{l|}{\,\,\,\,\,\,\,\,pthread}       & \textbf{92.9}  & \textbf{96.3}           & \textbf{94.5} & 61.4    & 80.6 & 92.6 & 86.2 & 28.7  & 61.5      & 59.3      & 60.4    & 46.0          & 35.7       & 55.6       & 43.5      & \textbf{20.0}            \\ \hline
\multicolumn{2}{l|}{RWIP}                          & \textbf{87.0}  & \textbf{95.2}  & \textbf{90.3} & 116.7   & 68.2           & 87.6          & 75.3          & 51.3           & 41.7      & 30.1      & 32.1    & 51.2          & 34.4       & 30.0       & 30.3      & \textbf{29.6}   \\
\multicolumn{2}{l|}{\,\,\,\,\,\,\,\,blink}         & \textbf{97.1}  & \textbf{97.1}  & \textbf{97.1} & 106.9   & 97.0           & 94.1          & 95.5          & 46.6          & 64.3      & 26.5      & 37.5    & 50.8          & 44.4       & 23.5       & 30.8      & \textbf{27.3}   \\
\multicolumn{2}{l|}{\,\,\,\,\,\,\,\,brake}         & \textbf{97.6} & \textbf{97.6}  & \textbf{97.6} & 140.4   & 65.3 & 80.0 & 71.9 & 61.6          & 45.5      & 12.5      & 19.6    & 47.2          & 33.3       & 14.7       & 20.4      & \textbf{27.4}   \\
\multicolumn{2}{l|}{\,\,\,\,\,\,\,\,i2c\_pca\_isa} & \textbf{90.2}            & \textbf{90.2}  & \textbf{90.2} & 120.2   & 70.0           & 80.3 & 74.8          & 48.5 & 50.0       & 57.4       & 53.4     & 49.6          & 48.6        & 57.4        & 52.6       & \textbf{26.8}   \\
\multicolumn{2}{l|}{\,\,\,\,\,\,\,\,i8xx\_tco}     & \textbf{62.5}           & \textbf{100.0} & \textbf{76.9} & 49.9    & 29.4          & \textbf{100.0}         & 45.5          & \textbf{29.6}  & 25.0       & 20.0      & 22.2    & 53.1          & 6.3        & 20.0        & 9.5       & 35.0           \\
\multicolumn{2}{l|}{\,\,\,\,\,\,\,\,logger}        & \textbf{85.7}  & \textbf{100.0} & \textbf{92.3} & 102.4   & 37.5           & \textbf{100.0}          & 54.5          & 46.5          & 20.0      & 50.0      & 28.6    & 46.6          & 15.8       & 50.0       & 24.0      & \textbf{27.8}   \\
\multicolumn{2}{l|}{\,\,\,\,\,\,\,\,wdt\_pci}      & \textbf{66.1}           & \textbf{97.5}  & \textbf{78.8} & 180.2   & 53.4           & 97.5          & 69.0          & 75.0          & 11.5      & 7.5       & 9.1    & 59.7          & 9.8       & 12.5        & 11.0      & \textbf{33.2}   \\ \hline
\multicolumn{2}{c|}{Total/Average}                 & \textbf{91.0}  & \textbf{96.4}  & \textbf{93.6} & 37.0    & 71.8           & 91.5          & 80.5          & 27.0  & 50.5      & 47.7      & 49.0    & 35.6          & 39.6       & 46.4       & 42.7      & \textbf{22.1}           \\ \hline
\end{tabular}

\begin{tablenotes}
\item [*] PA refers to the \emph{Plan Agent}. JA refers to the \emph{Judge agent}. \textbf{Bold numbers} indicate the best results.
\end{tablenotes}
\end{threeparttable}
}
\label{tab:ablation}
\end{table*}

To understand the contribution of each component in \tool, we conduct an ablation study, as shown in Table~\ref{tab:ablation}. As \emph{Expert Agent} is the core component responsible for detecting atomicity violations, our ablation study focuses on evaluating the impact of removing the \textit{Plan Agent} and \emph{Judge Agent}.

The removal of \textit{Plan Agent} (scheme B and C) results in a substantial degradation in \tool's performance ---  the number of detected defects is reduced by half and the false positive rate increases from 9.0\% to 49.5-60.4\%. This sharp effectiveness decline highlights the critical role of the \textit{Plan Agent} in enhancing LLM capability by concentrating its attention on essential code regions. Additionally, it reduces the execution time by 58\% through minimizing input tokens for LLM inference and eliminating unnecessary LLM invocations for non-existent atomicity violation patterns.

\begin{figure}
    \centering
    \includegraphics[width=\linewidth]{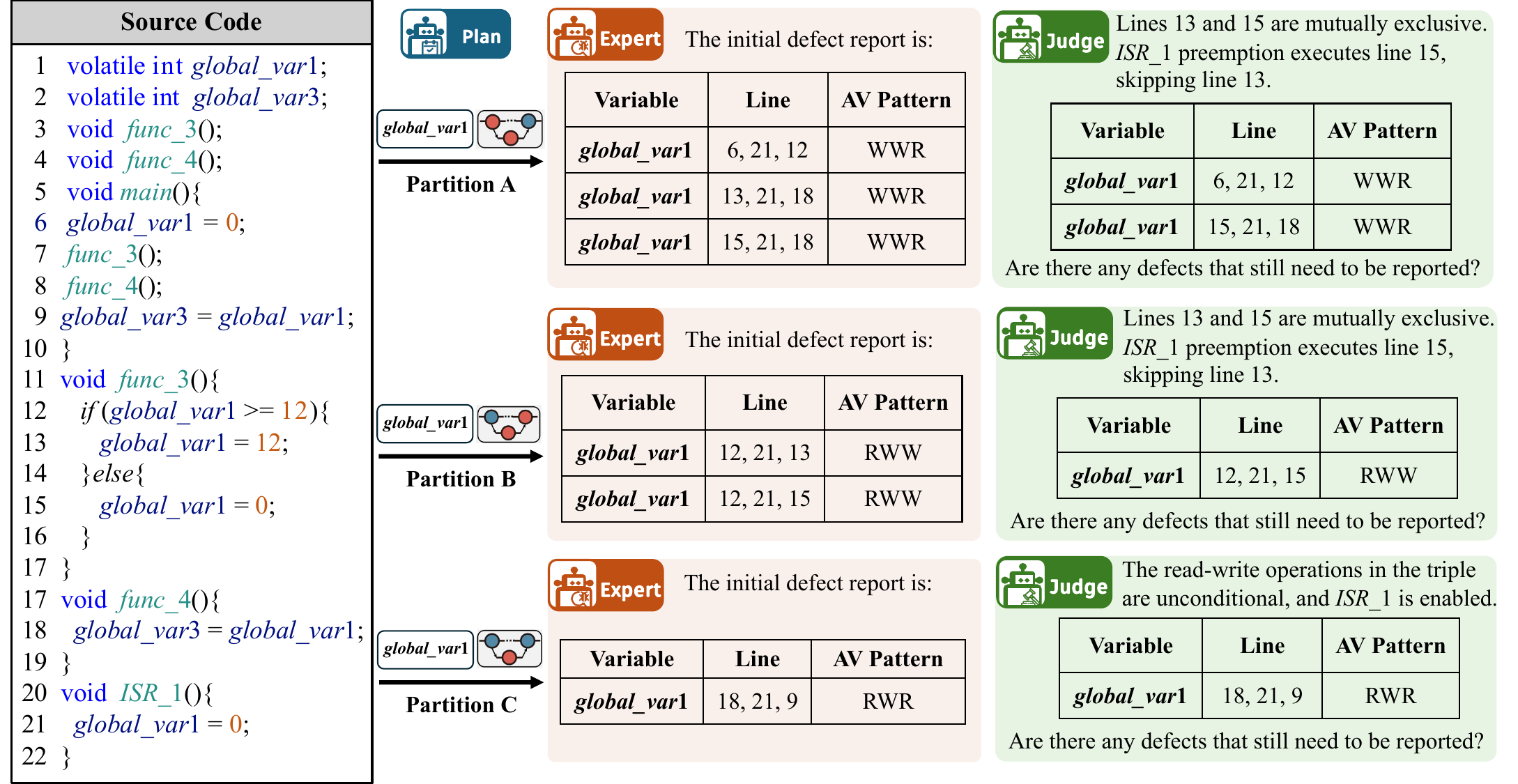}
    \caption{\tool @ Ex22 from \racebench.
    }
    \label{fig:RQ2-example}
\end{figure}

The results of scheme A demonstrate that the \textit{Judge Agent} helps reduce incorrect defect reports generated by the \textit{Expert Agent} by 26.7\%. Furthermore, it improves detection capability by identifying 5.4\% more atomicity violations through corrections and invoking next-round detection. 
To further understand the effect of \textit{Expert Agent}, consider Figure~\ref{fig:RQ2-example}. The \emph{Plan Agent} invokes static analysis tools to partition the analysis into three Expert-Judge tasks, \emph{Expert Agent} generates initial defect reports identifying potential atomicity violations, and \emph{Judge Agent} filters out two invalid defect reports caused by unsatisfiable branch conditions.

\begin{tcolorbox}[size=title,colframe=white,width=1\linewidth,colback=gray!20]
\textbf{Summary}: 
Every component of \tool contributes to its overall performance. \textit{Expert Agent} serves as the backbone of atomicity violation detection, and \textit{Plan Agent} contributes the most besides it. \textit{Judge Agent} further refines detection accuracy.
\end{tcolorbox}

\subsection{Impact of LLMs}
\label{sec:llm_impact}

\begin{table*}[]
\caption{Overall performance of \tool with different LLM settings across 3 benchmarks.}
\resizebox{\linewidth}{!}{
\small
\begin{threeparttable}
\begin{tabular}{c|c|c c c c|cccc|cccc|cccc}
\hline
\multirow{2}{*}{\textbf{Models}} & \multirow{2}{*}{\textbf{Size}} & \multicolumn{4}{c|}{\textbf{Racebench 2.1}} & \multicolumn{4}{c|}{\textbf{SV-COMP}} & \multicolumn{4}{c|}{\textbf{RWIP}} & \multicolumn{4}{c}{\textbf{Average}} \\
\cline{3-6} \cline{7-10} \cline{11-14} \cline{15-18}
& & Pre & Rec & F1 & Time(s) & Pre & Rec & F1 & Time(s) & Pre & Rec & F1 & Time(s) & Pre & Rec & F1 & Time(s) \\ \hline
\textit{\textbf{Closed-source LLMs}} & & & & & & & & & & & & & & & & & \\
Claude-Sonnet-4 (default) \includegraphics[width=1.2em]{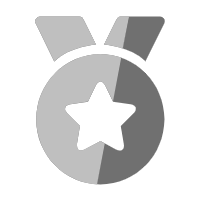} & - & 91.7& 97.2& 94.0& 45.0& 93.5& 96.3& 95.3& 33.9& 87.0& 95.2& 91.0& 116.7& 91.0& 96.4& 93.6& 37.0\\
Claude-3.5-Sonnet & - & 83.1& 94.7& 88.5& 26.3 & 90.0& 95.9& 93.3& 22.1& 81.2& 86.0& 83.6& 68.7& 85.1& 91.0& 87.9& 21.6\\
Gemini-2.5-Pro \includegraphics[width=1.2em]{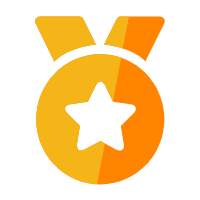} & - & \textbf{95.0} & \textbf{100.0}& \textbf{97.4}& 129.0& \textbf{96.7}& \textbf{99.3}& \textbf{98.0}& 96.6& \textbf{89.3}& \textbf{98.4}& \textbf{93.6}& 310.4& \textbf{92.9}& \textbf{99.0}& \textbf{95.8}& 131.0\\
Qwen-Max \includegraphics[width=1.2em]{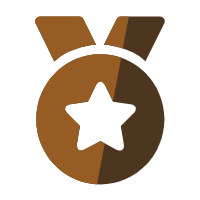} & - & 80.6& 94.7& 87.1& 80.3 & 89.4& \textbf{99.3}& 94.1& 58.9& 85.7& 80.6& 83.1& 277.9& 86.3& 89.7& 88.0& 91.6\\ \hline
\textit{\textbf{Open-source LLMs}} & & & & & & & & & & & & & & & & & \\
Kimi-K2 & 1T & 77.9 & 93.0 & 84.8 & 56.6 & 86.1& 97.9& 91.6& 46.5& 80.0& 89.8& 85.0& 207.4& 82.3& 93.3& 87.4& 69.2\\
DeepSeek-V3 & 236B & 73.2& 91.2& 81.3& \textbf{14.4} & 86.3& 86.2& 86.2& \textbf{11.0}& 72.0& 94.1& 81.6& 35.7& 77.5& 90.7& 83.6& \textbf{14.9}\\
LongCat-Flash-Chat & 560B & 88.3& 93.0& 90.6& 18.0 & 92.8& 95.2& 94.0& 19.9& 80.8& 83.9& 82.3& \textbf{35.1}& 86.4& 89.4& 87.9& 21.4\\
\hline
\end{tabular}
\begin{tablenotes}
\item [*] "Pre(\%)" refers to Precision, "Rec(\%)" refers to Recall, and "F1(\%)" refers to the F1-score. \textbf{Bold numbers} indicate the best results.
\end{tablenotes}
\end{threeparttable}
\label{tab:differ-LLM}
}
\end{table*}
To evaluate the impact of different LLMs on \tool's effectiveness, we replace the incorporated LLM with several advanced models from different LLM families and repeat the experiments described in Section~\ref{sec:eval_rq1}. 
The results are summarized in Table~\ref{tab:differ-LLM}.

Overall, the choice of LLM has a relatively minor impact on \tool's performance. This finding demonstrates the robustness and adaptability of \tool in leveraging state-of-the-art LLMs for atomicity violation detection. By effectively utilizing the reasoning and comprehension capabilities of different models, \tool consistently achieves high accuracy across LLM settings.

Claude-Sonnet-4~\cite{claude-4}, the default LLM setting of \tool, maintains competitive performance with an average recall of 96.4\% and F1-score of 93.6\% across all benchmarks. However, Gemini-2.5-Pro~\cite{comanici2025gemini} achieves the highest average precision (92.9\%), recall (99.0\%), and F1-score (95.8\%) across all benchmarks at the cost of five times longer execution time, due to its exceptional proficiency in code comprehension and defect identification. These results confirm that \tool's multi-agent architecture effectively harnesses the strengths of different LLMs while maintaining consistent high performance across various model choices. 

Generally, closed-source LLMs demonstrate superior performance compared to the open-source ones. LongCat-Flash-Chat~\cite{longcatflashtechnicalreport}, as an open-source model, demonstrates exceptional performance with an average F1-score of 87.9\%, matching the performance of closed-source models Claude-3.5-Sonnet (87.9\%) and Qwen-Max~\cite{qwen25} (88.0\%), while offering significant advantages in execution efficiency. With an average runtime of only 21.4 seconds, it is much faster than Qwen-Max (91.6s) and close to Claude-3.5-Sonnet (21.6s).

\section{Threat to Validity}
\emph{Internal threats to validity.}
\tool relies on static analysis to extract critical code snippets, which may fail to capture operations within third-party libraries or complex program dependencies. In addition, LLMs inevitably make mistakes in detecting atomicity violations, and which may not be eliminated through the limited number of iterations of multi-agent conversation. 

\emph{External threats to validity.}
Our evaluation is conducted on four C/C++ benchmark suites, which may not represent all interrupt-driven applications. The evaluation results may not generalize to programs with different interrupt mechanisms, hardware platforms, or software architectures.
\tool is only evaluated on 7 LLM settings. It may have different performance with other LLMs.

\section{Related Work}

\sloppy

\subsection{Atomicity Violation Detection}
Detecting {atomicity violations} in interrupt-driven software has gained much attention in recent years, aiming to identify improper operation sequences on shared resources when interrupts preempt critical sections.

One line of work adopts static code analysis for this task. 
IntAtom~\cite{li2022precise} incorporates a Staged Path Pruning method, iteratively discarding irrelevant execution paths in interrupt-driven programs.
CPA4AV~\cite{yu2023detecting} selects interrupt points and delaying ISR triggers, in order to effectively reduce reliance on complex scheduling and thus enhance detection accuracy.
SpecChecker-ISA~\cite{wang2022specchecker} leverages abstract interpretation and numeric modeling to accurately locate shared data access points in interrupt-driven software, thus reducing false positives. 

Another line of work incorporates run-time information to enhance detection effectiveness. 
CTrigger~\cite{5740930} identifies feasible non-serializable alternate executions by tracking and analyzing the program. 
A-VIO~\cite{lu2006avio} infers the programmer's expectations regarding atomicity assumptions and detects violations at runtime, capturing a broader range of atomicity violations. 
Rchecker~\cite{feng2020rchecker} integrates static and dynamic analysis, utilizing control path analysis and the C-bounded Model Checker, to detect data races in interrupt-driven embedded software. 

However, these methods face scalability challenges due to state-space explosion, limited domain-specific reasoning, and reliance on simplistic pattern recognition. To address these limitations, we propose \tool, a hybrid framework integrating the reliability of precise static analysis with the comprehension capability of domain-aware LLM agents. \tool outperforms them in terms of effectiveness and scalability.

\subsection{Integration of LLMs with Static Analysis}
The integration of large language models (LLMs) with static analysis has recently emerged as a promising approach to enhance the precision and efficiency of program analysis~\cite{wang2025contemporary}. 

For error-specification inference, Chapman et al.~\cite{10.1145/3652588.3663317} interleave calls to the static analyzer and queries to the LLM, inferring the set of values returned by each function upon error. 
For loop invariants identification, E\&V~\cite{hao2023evpromptinglargelanguage} employs LLMs to simulate the execution of pseudo-code, without the need of an external oracle.
For resource leak detection, INFERROI~\cite{wang2023boosting} utilizes LLM to directly infer resource-oriented intentions, followed by a two-stage static analysis approach.
For detecting use-before-initialization (UBI) errors in the Linux kernel,  LLift~\cite{li2024enhancing} enhances path analysis using post-constraint guidance obtained by the LLM.
For data flow analysis, LLMDFA~\cite{wang2024llmdfa} leverages LLMs to synthesize code that outsources delicate reasoning to external expert tools and summarizes data flow facts in individual functions.

These studies underscore the significant potential of integrating static analysis techniques with LLMs. However, they target different program analysis tasks from our work. In addition, they utilize large LLMs in a straightforward manner, such as direct prompting~\cite{10.1145/3652588.3663317, hao2023evpromptinglargelanguage, li2024enhancing} and chain-of-thought~\cite{ wang2023boosting,wang2024llmdfa} techniques. 
In contrast, our approach adopts a multi-agent framework that maximizes LLMs' code comprehension capabilities through structured collaboration and static analysis synergy.

\subsection{Tackling defects with LLM Agents}
Agent-based LLM frameworks have gained significant attention in tasks of vulnerability detection, automated fixing, and code review~\cite{liu2024large}. Prior work leverages LLMs to capture context semantics, often augmented with specialized knowledge bases or external tools for more targeted analysis.

Vul-RAG~\cite{du2024vul} is a retrieval-augmented generation (RAG) technique that leverages a knowledge base of CVE instances to detect and fix security flaws.
CODEAGENT~\cite{zhang2024codeagent} integrates external tools for large-scale code generation and testing, providing an end-to-end software development pipeline.
CODER~\cite{chen2024coder} utilizes a pre-defined task graph to guide multiple specialized agents to resolve general defects and add new software features.
MarsCode~\cite{liu2024marscode} incorporates LLMs for automated code repairing, following the process of planning, reproduction, fault localization, patching, and validation.
CodeAgent~\cite{tang2024codeagent} and Mao et al.~\cite{mao2024multi} propose multi-agent systems to replicate real-life code review processes, where each LLM agent plays a specific role.
Another group of work~\cite{islam2024mapcoder, nguyen2024agilecoder, zhang2024autocoderover, qian2024scaling} utilizes multi-agent collaboration to tackle general software engineering tasks.

Despite their advancements, they focus on general-purpose software, lacking domain knowledge for domain-specific programs. In addition, they are under-validated in safety-critical environments. In contrast, our \tool addresses these limitations by integrating static analysis with several specialized agents, enabling precise concurrency-focused analysis.

\section{Conclusion}

Atomicity violations threaten the correctness and safety of interrupt-driven programs, by disrupting the intended operation sequence on shared resources. It is challenging to detect them due to program state space reasoning, application-level code context, and complex domain-specific knowledge.  

This paper presents \tool, a multi-agent framework that detects atomicity violation in real-world interrupt-driven programs. Its plan agent orchestrates four static analysis tools to extract key information and generate code summary. \tool then initializes several \emph{Expert-Judge} agent pairs to detect and validate different patterns of atomicity violation, through an iterative manner. 
\tool is evaluated with a variety of real-world programs of different scales, and out-performs existing approaches.


  \newpage

\bibliographystyle{ACM-Reference-Format}
\bibliography{citation.bib}


\begin{thebibliography}{78}


\ifx \showCODEN    \undefined \def \showCODEN     #1{\unskip}     \fi
\ifx \showDOI      \undefined \def \showDOI       #1{#1}\fi
\ifx \showISBNx    \undefined \def \showISBNx     #1{\unskip}     \fi
\ifx \showISBNxiii \undefined \def \showISBNxiii  #1{\unskip}     \fi
\ifx \showISSN     \undefined \def \showISSN      #1{\unskip}     \fi
\ifx \showLCCN     \undefined \def \showLCCN      #1{\unskip}     \fi
\ifx \shownote     \undefined \def \shownote      #1{#1}          \fi
\ifx \showarticletitle \undefined \def \showarticletitle #1{#1}   \fi
\ifx \showURL      \undefined \def \showURL       {\relax}        \fi
\providecommand\bibfield[2]{#2}
\providecommand\bibinfo[2]{#2}
\providecommand\natexlab[1]{#1}
\providecommand\showeprint[2][]{arXiv:#2}

\bibitem[Sung et~al\mbox{.}(2017)]%
        {8115634}
\bibfield{author}{\bibinfo{person}{Chungha Sung}, \bibinfo{person}{Markus Kusano}, {and} \bibinfo{person}{Chao Wang}.} \bibinfo{year}{2017}\natexlab{}.
\newblock \showarticletitle{Modular verification of interrupt-driven software}. In \bibinfo{booktitle}{\emph{2017 32nd IEEE/ACM International Conference on Automated Software Engineering (ASE)}}. \bibinfo{pages}{206--216}.
\newblock
\urldef\tempurl%
\url{https://doi.org/10.1109/ASE.2017.8115634}
\showDOI{\tempurl}


\bibitem[Chen et~al\mbox{.}(2011)]%
        {6004502}
\bibfield{author}{\bibinfo{person}{Rui Chen}, \bibinfo{person}{Xiangying Guo}, \bibinfo{person}{Yonghao Duan}, \bibinfo{person}{Bin Gu}, {and} \bibinfo{person}{Mengfei Yang}.} \bibinfo{year}{2011}\natexlab{}.
\newblock \showarticletitle{Static Data Race Detection for Interrupt-Driven Embedded Software}. In \bibinfo{booktitle}{\emph{2011 Fifth International Conference on Secure Software Integration and Reliability Improvement - Companion}}. \bibinfo{pages}{47--52}.
\newblock
\urldef\tempurl%
\url{https://doi.org/10.1109/SSIRI-C.2011.18}
\showDOI{\tempurl}


\bibitem[Parmer and West(2008)]%
        {parmer2008predictable}
\bibfield{author}{\bibinfo{person}{Gabriel Parmer} {and} \bibinfo{person}{Richard West}.} \bibinfo{year}{2008}\natexlab{}.
\newblock \showarticletitle{Predictable interrupt management and scheduling in the Composite component-based system}. In \bibinfo{booktitle}{\emph{2008 Real-Time Systems Symposium}}. IEEE, \bibinfo{pages}{232--243}.
\newblock


\bibitem[Leveson and Harvey(1983)]%
        {leveson1983analyzing}
\bibfield{author}{\bibinfo{person}{Nancy~G. Leveson} {and} \bibinfo{person}{Peter~R. Harvey}.} \bibinfo{year}{1983}\natexlab{}.
\newblock \showarticletitle{Analyzing software safety}.
\newblock \bibinfo{journal}{\emph{IEEE Transactions on Software Engineering}} \bibinfo{number}{5} (\bibinfo{year}{1983}), \bibinfo{pages}{569--579}.
\newblock


\bibitem[Lu et~al\mbox{.}(2011)]%
        {lu2011finding}
\bibfield{author}{\bibinfo{person}{Shan Lu}, \bibinfo{person}{Soyeon Park}, {and} \bibinfo{person}{Yuanyuan Zhou}.} \bibinfo{year}{2011}\natexlab{}.
\newblock \showarticletitle{Finding atomicity-violation bugs through unserializable interleaving testing}.
\newblock \bibinfo{journal}{\emph{IEEE Transactions on Software Engineering}} \bibinfo{volume}{38}, \bibinfo{number}{4} (\bibinfo{year}{2011}), \bibinfo{pages}{844--860}.
\newblock


\bibitem[Jin et~al\mbox{.}(2011)]%
        {jin2011automated}
\bibfield{author}{\bibinfo{person}{Guoliang Jin}, \bibinfo{person}{Linhai Song}, \bibinfo{person}{Wei Zhang}, \bibinfo{person}{Shan Lu}, {and} \bibinfo{person}{Ben Liblit}.} \bibinfo{year}{2011}\natexlab{}.
\newblock \showarticletitle{Automated atomicity-violation fixing}. In \bibinfo{booktitle}{\emph{Proceedings of the 32nd ACM SIGPLAN conference on Programming language design and implementation}}. \bibinfo{pages}{389--400}.
\newblock


\bibitem[Wu et~al\mbox{.}(2015)]%
        {7318260}
\bibfield{author}{\bibinfo{person}{Xueguang Wu}, \bibinfo{person}{Liqian Chen}, \bibinfo{person}{Antoine Mine}, \bibinfo{person}{Wei Dong}, {and} \bibinfo{person}{Ji Wang}.} \bibinfo{year}{2015}\natexlab{}.
\newblock \showarticletitle{Numerical static analysis of interrupt-driven programs via sequentialization}. In \bibinfo{booktitle}{\emph{2015 International Conference on Embedded Software (EMSOFT)}}. \bibinfo{pages}{55--64}.
\newblock
\urldef\tempurl%
\url{https://doi.org/10.1109/EMSOFT.2015.7318260}
\showDOI{\tempurl}


\bibitem[Regehr(2005)]%
        {regehr2005random}
\bibfield{author}{\bibinfo{person}{John Regehr}.} \bibinfo{year}{2005}\natexlab{}.
\newblock \showarticletitle{Random testing of interrupt-driven software}. In \bibinfo{booktitle}{\emph{Proceedings of the 5th ACM International Conference on Embedded software}}. \bibinfo{pages}{290--298}.
\newblock


\bibitem[Park et~al\mbox{.}(2009)]%
        {park2009ctrigger}
\bibfield{author}{\bibinfo{person}{Soyeon Park}, \bibinfo{person}{Shan Lu}, {and} \bibinfo{person}{Yuanyuan Zhou}.} \bibinfo{year}{2009}\natexlab{}.
\newblock \showarticletitle{CTrigger: exposing atomicity violation bugs from their hiding places}. In \bibinfo{booktitle}{\emph{Proceedings of the 14th international conference on Architectural support for programming languages and operating systems}}. \bibinfo{pages}{25--36}.
\newblock


\bibitem[Gu et~al\mbox{.}(2015)]%
        {gu2015change}
\bibfield{author}{\bibinfo{person}{Rui Gu}, \bibinfo{person}{Guoliang Jin}, \bibinfo{person}{Linhai Song}, \bibinfo{person}{Linjie Zhu}, {and} \bibinfo{person}{Shan Lu}.} \bibinfo{year}{2015}\natexlab{}.
\newblock \showarticletitle{What change history tells us about thread synchronization}. In \bibinfo{booktitle}{\emph{Proceedings of the 2015 10th Joint Meeting on Foundations of Software Engineering}}. \bibinfo{pages}{426--438}.
\newblock


\bibitem[Halstead~Jr(1985)]%
        {halstead1985multilisp}
\bibfield{author}{\bibinfo{person}{Robert~H Halstead~Jr}.} \bibinfo{year}{1985}\natexlab{}.
\newblock \showarticletitle{Multilisp: A language for concurrent symbolic computation}.
\newblock \bibinfo{journal}{\emph{ACM Transactions on Programming Languages and Systems (TOPLAS)}} \bibinfo{volume}{7}, \bibinfo{number}{4} (\bibinfo{year}{1985}), \bibinfo{pages}{501--538}.
\newblock


\bibitem[Lu et~al\mbox{.}(2008)]%
        {lu2008learning}
\bibfield{author}{\bibinfo{person}{Shan Lu}, \bibinfo{person}{Soyeon Park}, \bibinfo{person}{Eunsoo Seo}, {and} \bibinfo{person}{Yuanyuan Zhou}.} \bibinfo{year}{2008}\natexlab{}.
\newblock \showarticletitle{Learning from mistakes: a comprehensive study on real world concurrency bug characteristics}. In \bibinfo{booktitle}{\emph{Proceedings of the 13th international conference on Architectural support for programming languages and operating systems}}. \bibinfo{pages}{329--339}.
\newblock


\bibitem[Kroening et~al\mbox{.}(2015)]%
        {7092387}
\bibfield{author}{\bibinfo{person}{Daniel Kroening}, \bibinfo{person}{Lihao Liang}, \bibinfo{person}{Tom Melham}, \bibinfo{person}{Peter Schrammel}, {and} \bibinfo{person}{Michael Tautschnig}.} \bibinfo{year}{2015}\natexlab{}.
\newblock \showarticletitle{Effective verification of low-level software with nested interrupts}. In \bibinfo{booktitle}{\emph{2015 Design, Automation \& Test in Europe Conference \& Exhibition (DATE)}}. \bibinfo{pages}{229--234}.
\newblock
\urldef\tempurl%
\url{https://doi.org/10.7873/DATE.2015.0360}
\showDOI{\tempurl}


\bibitem[Pradel et~al\mbox{.}(2014)]%
        {pradel2014performance}
\bibfield{author}{\bibinfo{person}{Michael Pradel}, \bibinfo{person}{Markus Huggler}, {and} \bibinfo{person}{Thomas~R Gross}.} \bibinfo{year}{2014}\natexlab{}.
\newblock \showarticletitle{Performance regression testing of concurrent classes}. In \bibinfo{booktitle}{\emph{Proceedings of the 2014 International Symposium on Software Testing and Analysis}}. \bibinfo{pages}{13--25}.
\newblock


\bibitem[Tu et~al\mbox{.}(2019)]%
        {tu2019understanding}
\bibfield{author}{\bibinfo{person}{Tengfei Tu}, \bibinfo{person}{Xiaoyu Liu}, \bibinfo{person}{Linhai Song}, {and} \bibinfo{person}{Yiying Zhang}.} \bibinfo{year}{2019}\natexlab{}.
\newblock \showarticletitle{Understanding real-world concurrency bugs in go}. In \bibinfo{booktitle}{\emph{Proceedings of the twenty-fourth international conference on architectural support for programming languages and operating systems}}. \bibinfo{pages}{865--878}.
\newblock


\bibitem[rac(2019)]%
        {racebench}
 \bibinfo{year}{2019}\natexlab{}.
\newblock \bibinfo{title}{Racebench 2.1}.
\newblock \bibinfo{howpublished}{\url{https://github.com/chenruibuaa/racebench/tree/master}}.
\newblock


\bibitem[Lu et~al\mbox{.}(2007)]%
        {lu2007muvi}
\bibfield{author}{\bibinfo{person}{Shan Lu}, \bibinfo{person}{Soyeon Park}, \bibinfo{person}{Chongfeng Hu}, \bibinfo{person}{Xiao Ma}, \bibinfo{person}{Weihang Jiang}, \bibinfo{person}{Zhenmin Li}, \bibinfo{person}{Raluca~A Popa}, {and} \bibinfo{person}{Yuanyuan Zhou}.} \bibinfo{year}{2007}\natexlab{}.
\newblock \showarticletitle{MUVI: Automatically inferring multi-variable access correlations and detecting related semantic and concurrency bugs}. In \bibinfo{booktitle}{\emph{Proceedings of twenty-first ACM SIGOPS symposium on Operating systems principles}}. \bibinfo{pages}{103--116}.
\newblock


\bibitem[Li et~al\mbox{.}(2022)]%
        {li2022precise}
\bibfield{author}{\bibinfo{person}{Chao Li}, \bibinfo{person}{Rui Chen}, \bibinfo{person}{Boxiang Wang}, \bibinfo{person}{Tingting Yu}, \bibinfo{person}{Dongdong Gao}, {and} \bibinfo{person}{Mengfei Yang}.} \bibinfo{year}{2022}\natexlab{}.
\newblock \showarticletitle{Precise and efficient atomicity violation detection for interrupt-driven programs via staged path pruning}. In \bibinfo{booktitle}{\emph{Proceedings of the 31st ACM SIGSOFT International Symposium on Software Testing and Analysis}}. \bibinfo{pages}{506--518}.
\newblock


\bibitem[Mogul and Ramakrishnan(1997)]%
        {mogul1997eliminating}
\bibfield{author}{\bibinfo{person}{Jeffrey~C Mogul} {and} \bibinfo{person}{Kadangode~K Ramakrishnan}.} \bibinfo{year}{1997}\natexlab{}.
\newblock \showarticletitle{Eliminating receive livelock in an interrupt-driven kernel}.
\newblock \bibinfo{journal}{\emph{ACM Transactions on Computer Systems}} \bibinfo{volume}{15}, \bibinfo{number}{3} (\bibinfo{year}{1997}), \bibinfo{pages}{217--252}.
\newblock


\bibitem[Lu et~al\mbox{.}(2006)]%
        {lu2006avio}
\bibfield{author}{\bibinfo{person}{Shan Lu}, \bibinfo{person}{Joseph Tucek}, \bibinfo{person}{Feng Qin}, {and} \bibinfo{person}{Yuanyuan Zhou}.} \bibinfo{year}{2006}\natexlab{}.
\newblock \showarticletitle{AVIO: detecting atomicity violations via access interleaving invariants}.
\newblock \bibinfo{journal}{\emph{ACM SIGOPS Operating Systems Review}} \bibinfo{volume}{40}, \bibinfo{number}{5} (\bibinfo{year}{2006}), \bibinfo{pages}{37--48}.
\newblock


\bibitem[Li et~al\mbox{.}(2023)]%
        {10.1145/3597926.3598140}
\bibfield{author}{\bibinfo{person}{Chao Li}, \bibinfo{person}{Rui Chen}, \bibinfo{person}{Boxiang Wang}, \bibinfo{person}{Zhixuan Wang}, \bibinfo{person}{Tingting Yu}, \bibinfo{person}{Yunsong Jiang}, \bibinfo{person}{Bin Gu}, {and} \bibinfo{person}{Mengfei Yang}.} \bibinfo{year}{2023}\natexlab{}.
\newblock \showarticletitle{An Empirical Study on Concurrency Bugs in Interrupt-Driven Embedded Software} \emph{(\bibinfo{series}{ISSTA 2023})}.
\newblock
\showISBNx{9798400702211}
\urldef\tempurl%
\url{https://doi.org/10.1145/3597926.3598140}
\showDOI{\tempurl}


\bibitem[Feng et~al\mbox{.}(2020)]%
        {feng2020rchecker}
\bibfield{author}{\bibinfo{person}{Haining Feng}, \bibinfo{person}{Liangze Yin}, \bibinfo{person}{Wenfeng Lin}, \bibinfo{person}{Xudong Zhao}, {and} \bibinfo{person}{Wei Dong}.} \bibinfo{year}{2020}\natexlab{}.
\newblock \showarticletitle{Rchecker: A cbmc-based data race detector for interrupt-driven programs}. In \bibinfo{booktitle}{\emph{2020 IEEE 20th International Conference on Software Quality, Reliability and Security Companion (QRS-C)}}. IEEE, \bibinfo{pages}{465--471}.
\newblock


\bibitem[Pan et~al\mbox{.}(2019)]%
        {8812085}
\bibfield{author}{\bibinfo{person}{Minxue Pan}, \bibinfo{person}{Shouyu Chen}, \bibinfo{person}{Yu Pei}, \bibinfo{person}{Tian Zhang}, {and} \bibinfo{person}{Xuandong Li}.} \bibinfo{year}{2019}\natexlab{}.
\newblock \showarticletitle{Easy Modelling and Verification of Unpredictable and Preemptive Interrupt-Driven Systems}. In \bibinfo{booktitle}{\emph{2019 IEEE/ACM 41st International Conference on Software Engineering (ICSE)}}. \bibinfo{pages}{212--222}.
\newblock
\urldef\tempurl%
\url{https://doi.org/10.1109/ICSE.2019.00037}
\showDOI{\tempurl}


\bibitem[Yu et~al\mbox{.}(2023)]%
        {yu2023detecting}
\bibfield{author}{\bibinfo{person}{Bin Yu}, \bibinfo{person}{Cong Tian}, \bibinfo{person}{Hengrui Xing}, \bibinfo{person}{Zuchao Yang}, \bibinfo{person}{Jie Su}, \bibinfo{person}{Xu Lu}, \bibinfo{person}{Jiyu Yang}, \bibinfo{person}{Liang Zhao}, \bibinfo{person}{Xiaofeng Li}, {and} \bibinfo{person}{Zhenhua Duan}.} \bibinfo{year}{2023}\natexlab{}.
\newblock \showarticletitle{Detecting Atomicity Violations in Interrupt-Driven Programs via Interruption Points Selecting and Delayed ISR-Triggering}. In \bibinfo{booktitle}{\emph{Proceedings of the 31st ACM Joint European Software Engineering Conference and Symposium on the Foundations of Software Engineering}}. \bibinfo{pages}{1153--1164}.
\newblock


\bibitem[Hou et~al\mbox{.}(2024)]%
        {10.1145/3695988}
\bibfield{author}{\bibinfo{person}{Xinyi Hou}, \bibinfo{person}{Yanjie Zhao}, \bibinfo{person}{Yue Liu}, \bibinfo{person}{Zhou Yang}, \bibinfo{person}{Kailong Wang}, \bibinfo{person}{Li Li}, \bibinfo{person}{Xiapu Luo}, \bibinfo{person}{David Lo}, \bibinfo{person}{John Grundy}, {and} \bibinfo{person}{Haoyu Wang}.} \bibinfo{year}{2024}\natexlab{}.
\newblock \showarticletitle{Large Language Models for Software Engineering: A Systematic Literature Review}.
\newblock \bibinfo{journal}{\emph{ACM Trans. Softw. Eng. Methodol.}} \bibinfo{volume}{33}, \bibinfo{number}{8}, Article \bibinfo{articleno}{220} (\bibinfo{date}{Dec.} \bibinfo{year}{2024}), \bibinfo{numpages}{79}~pages.
\newblock
\showISSN{1049-331X}
\urldef\tempurl%
\url{https://doi.org/10.1145/3695988}
\showDOI{\tempurl}


\bibitem[Zhao et~al\mbox{.}(2024)]%
        {zhao2024enhancing}
\bibfield{author}{\bibinfo{person}{Jiuang Zhao}, \bibinfo{person}{Donghao Yang}, \bibinfo{person}{Li Zhang}, \bibinfo{person}{Xiaoli Lian}, \bibinfo{person}{Zitian Yang}, {and} \bibinfo{person}{Fang Liu}.} \bibinfo{year}{2024}\natexlab{}.
\newblock \showarticletitle{Enhancing automated program repair with solution design}. In \bibinfo{booktitle}{\emph{Proceedings of the 39th IEEE/ACM International Conference on Automated Software Engineering}}. \bibinfo{pages}{1706--1718}.
\newblock


\bibitem[Hou et~al\mbox{.}(2024)]%
        {hou2024large}
\bibfield{author}{\bibinfo{person}{Xinyi Hou}, \bibinfo{person}{Yanjie Zhao}, \bibinfo{person}{Yue Liu}, \bibinfo{person}{Zhou Yang}, \bibinfo{person}{Kailong Wang}, \bibinfo{person}{Li Li}, \bibinfo{person}{Xiapu Luo}, \bibinfo{person}{David Lo}, \bibinfo{person}{John Grundy}, {and} \bibinfo{person}{Haoyu Wang}.} \bibinfo{year}{2024}\natexlab{}.
\newblock \showarticletitle{Large language models for software engineering: A systematic literature review}.
\newblock \bibinfo{journal}{\emph{ACM Transactions on Software Engineering and Methodology}} \bibinfo{volume}{33}, \bibinfo{number}{8} (\bibinfo{year}{2024}), \bibinfo{pages}{1--79}.
\newblock


\bibitem[Tang et~al\mbox{.}(2024)]%
        {tang2024codeagent}
\bibfield{author}{\bibinfo{person}{Xunzhu Tang}, \bibinfo{person}{Kisub Kim}, \bibinfo{person}{Yewei Song}, \bibinfo{person}{Cedric Lothritz}, \bibinfo{person}{Bei Li}, \bibinfo{person}{Saad Ezzini}, \bibinfo{person}{Haoye Tian}, \bibinfo{person}{Jacques Klein}, {and} \bibinfo{person}{Tegawend{\'e} Bissyand{\'e}}.} \bibinfo{year}{2024}\natexlab{}.
\newblock \showarticletitle{CodeAgent: Autonomous Communicative Agents for Code Review}. In \bibinfo{booktitle}{\emph{Proceedings of the 2024 Conference on Empirical Methods in Natural Language Processing}}. \bibinfo{pages}{11279--11313}.
\newblock


\bibitem[Yildiz et~al\mbox{.}(2025)]%
        {yildiz2025benchmarking}
\bibfield{author}{\bibinfo{person}{Alperen Yildiz}, \bibinfo{person}{Sin~G Teo}, \bibinfo{person}{Yiling Lou}, \bibinfo{person}{Yebo Feng}, \bibinfo{person}{Chong Wang}, {and} \bibinfo{person}{Dinil~M Divakaran}.} \bibinfo{year}{2025}\natexlab{}.
\newblock \showarticletitle{Benchmarking LLMs and LLM-based Agents in Practical Vulnerability Detection for Code Repositories}.
\newblock \bibinfo{journal}{\emph{arXiv preprint arXiv:2503.03586}} (\bibinfo{year}{2025}).
\newblock


\bibitem[Xu et~al\mbox{.}(2022)]%
        {xu2022towards}
\bibfield{author}{\bibinfo{person}{Yinan Xu}, \bibinfo{person}{Zihao Yu}, \bibinfo{person}{Dan Tang}, \bibinfo{person}{Guokai Chen}, \bibinfo{person}{Lu Chen}, \bibinfo{person}{Lingrui Gou}, \bibinfo{person}{Yue Jin}, \bibinfo{person}{Qianruo Li}, \bibinfo{person}{Xin Li}, \bibinfo{person}{Zuojun Li}, {et~al\mbox{.}}} \bibinfo{year}{2022}\natexlab{}.
\newblock \showarticletitle{Towards developing high performance RISC-V processors using agile methodology}. In \bibinfo{booktitle}{\emph{2022 55th IEEE/ACM International Symposium on Microarchitecture (MICRO)}}. IEEE, \bibinfo{pages}{1178--1199}.
\newblock


\bibitem[Cai et~al\mbox{.}(2021)]%
        {cai2021canary}
\bibfield{author}{\bibinfo{person}{Yuandao Cai}, \bibinfo{person}{Peisen Yao}, {and} \bibinfo{person}{Charles Zhang}.} \bibinfo{year}{2021}\natexlab{}.
\newblock \showarticletitle{Canary: practical static detection of inter-thread value-flow bugs}. In \bibinfo{booktitle}{\emph{Proceedings of the 42nd ACM SIGPLAN International Conference on Programming Language Design and Implementation}}. \bibinfo{pages}{1126--1140}.
\newblock


\bibitem[Musuvathi et~al\mbox{.}(2002)]%
        {musuvathi2002cmc}
\bibfield{author}{\bibinfo{person}{Madanlal Musuvathi}, \bibinfo{person}{David~YW Park}, \bibinfo{person}{Andy Chou}, \bibinfo{person}{Dawson~R Engler}, {and} \bibinfo{person}{David~L Dill}.} \bibinfo{year}{2002}\natexlab{}.
\newblock \showarticletitle{CMC: A pragmatic approach to model checking real code}.
\newblock \bibinfo{journal}{\emph{ACM SIGOPS Operating Systems Review}} \bibinfo{volume}{36}, \bibinfo{number}{SI} (\bibinfo{year}{2002}), \bibinfo{pages}{75--88}.
\newblock


\bibitem[Yu et~al\mbox{.}(2023)]%
        {2022-20908}
\bibfield{author}{\bibinfo{person}{Tingting Yu}, \bibinfo{person}{Chao Li}, \bibinfo{person}{Boxiang Wang}, \bibinfo{person}{Rui Chen}, {and} \bibinfo{person}{Yunsong. Jiang}.} \bibinfo{year}{2023}\natexlab{}.
\newblock \showarticletitle{Atomicity Violation Detection for Interrupt-Driven Aerospace Embedded Software}.
\newblock \bibinfo{journal}{\emph{Journal of Computer Research and Development}} \bibinfo{volume}{60}, \bibinfo{number}{2} (\bibinfo{year}{2023}), \bibinfo{pages}{294--310}.
\newblock
\showISSN{1000-1239}
\urldef\tempurl%
\url{https://doi.org/10.7544/issn1000-1239.202220908}
\showDOI{\tempurl}


\bibitem[Zeng et~al\mbox{.}(2024)]%
        {zeng2024memorize}
\bibfield{author}{\bibinfo{person}{Zhiyuan Zeng}, \bibinfo{person}{Qipeng Guo}, \bibinfo{person}{Xiaoran Liu}, \bibinfo{person}{Zhangyue Yin}, \bibinfo{person}{Wentao Shu}, \bibinfo{person}{Mianqiu Huang}, \bibinfo{person}{Bo Wang}, \bibinfo{person}{Yunhua Zhou}, \bibinfo{person}{Linlin Li}, \bibinfo{person}{Qun Liu}, {et~al\mbox{.}}} \bibinfo{year}{2024}\natexlab{}.
\newblock \showarticletitle{Memorize step by step: Efficient long-context prefilling with incremental memory and decremental chunk}. In \bibinfo{booktitle}{\emph{Proceedings of the 2024 Conference on Empirical Methods in Natural Language Processing}}. \bibinfo{pages}{21021--21034}.
\newblock


\bibitem[Huang et~al\mbox{.}(2024)]%
        {huang2024recurrent}
\bibfield{author}{\bibinfo{person}{Chensen Huang}, \bibinfo{person}{Guibo Zhu}, \bibinfo{person}{Xuepeng Wang}, \bibinfo{person}{Yifei Luo}, \bibinfo{person}{Guojing Ge}, \bibinfo{person}{Haoran Chen}, \bibinfo{person}{Dong Yi}, {and} \bibinfo{person}{Jinqiao Wang}.} \bibinfo{year}{2024}\natexlab{}.
\newblock \showarticletitle{Recurrent context compression: Efficiently expanding the context window of llm}.
\newblock \bibinfo{journal}{\emph{arXiv preprint arXiv:2406.06110}} (\bibinfo{year}{2024}).
\newblock


\bibitem[Wang et~al\mbox{.}(2023)]%
        {wang2023does}
\bibfield{author}{\bibinfo{person}{Zhilong Wang}, \bibinfo{person}{Lan Zhang}, \bibinfo{person}{Chen Cao}, \bibinfo{person}{Nanqing Luo}, \bibinfo{person}{Xinzhi Luo}, {and} \bibinfo{person}{Peng Liu}.} \bibinfo{year}{2023}\natexlab{}.
\newblock \showarticletitle{How Does Naming Affect LLMs on Code Analysis Tasks?}
\newblock \bibinfo{journal}{\emph{arXiv preprint arXiv:2307.12488}} (\bibinfo{year}{2023}).
\newblock


\bibitem[Lehtinen et~al\mbox{.}(2024)]%
        {lehtinen2024let}
\bibfield{author}{\bibinfo{person}{Teemu Lehtinen}, \bibinfo{person}{Charles Koutcheme}, {and} \bibinfo{person}{Arto Hellas}.} \bibinfo{year}{2024}\natexlab{}.
\newblock \showarticletitle{Let's Ask AI About Their Programs: Exploring ChatGPT's Answers To Program Comprehension Questions}. In \bibinfo{booktitle}{\emph{Proceedings of the 46th International Conference on Software Engineering: Software Engineering Education and Training}}. \bibinfo{pages}{221--232}.
\newblock


\bibitem[Guo et~al\mbox{.}(2024)]%
        {guo2024largelanguagemodelbased}
\bibfield{author}{\bibinfo{person}{Taicheng Guo}, \bibinfo{person}{Xiuying Chen}, \bibinfo{person}{Yaqi Wang}, \bibinfo{person}{Ruidi Chang}, \bibinfo{person}{Shichao Pei}, \bibinfo{person}{Nitesh~V. Chawla}, \bibinfo{person}{Olaf Wiest}, {and} \bibinfo{person}{Xiangliang Zhang}.} \bibinfo{year}{2024}\natexlab{}.
\newblock \bibinfo{title}{Large Language Model based Multi-Agents: A Survey of Progress and Challenges}.
\newblock
\newblock
\showeprint[arxiv]{2402.01680}~[cs.CL]
\urldef\tempurl%
\url{https://arxiv.org/abs/2402.01680}
\showURL{%
\tempurl}


\bibitem[SVC(2022)]%
        {SVCOMP2022}
 \bibinfo{year}{2022}\natexlab{}.
\newblock \bibinfo{title}{SV-COMP 2022 Benchmark}.
\newblock \bibinfo{howpublished}{\url{https://github.com/BinYu-Xidian-University/CPA4AV/tree/main/SV-COMP-2022-Benchmark}}.
\newblock


\bibitem[int(2023)]%
        {interruptdriven}
 \bibinfo{year}{2023}\natexlab{}.
\newblock \bibinfo{title}{Real-world Interrupt-driven Dataset}.
\newblock \bibinfo{howpublished}{\url{https://github.com/BinYu-Xidian-University/CPA4AV/tree/main/Real-world-interrupt-driven}}.
\newblock


\bibitem[Pan et~al\mbox{.}(2019)]%
        {pan2019easy}
\bibfield{author}{\bibinfo{person}{Minxue Pan}, \bibinfo{person}{Shouyu Chen}, \bibinfo{person}{Yu Pei}, \bibinfo{person}{Tian Zhang}, {and} \bibinfo{person}{Xuandong Li}.} \bibinfo{year}{2019}\natexlab{}.
\newblock \showarticletitle{Easy modelling and verification of unpredictable and preemptive interrupt-driven systems}. In \bibinfo{booktitle}{\emph{2019 IEEE/ACM 41st International Conference on Software Engineering (ICSE)}}. IEEE, \bibinfo{pages}{212--222}.
\newblock


\bibitem[Schwarz et~al\mbox{.}(2011)]%
        {schwarz2011static}
\bibfield{author}{\bibinfo{person}{Martin~D Schwarz}, \bibinfo{person}{Helmut Seidl}, \bibinfo{person}{Vesal Vojdani}, \bibinfo{person}{Peter Lammich}, {and} \bibinfo{person}{Markus M{\"u}ller-Olm}.} \bibinfo{year}{2011}\natexlab{}.
\newblock \showarticletitle{Static analysis of interrupt-driven programs synchronized via the priority ceiling protocol}.
\newblock \bibinfo{journal}{\emph{ACM SIGPLAN Notices}} \bibinfo{volume}{46}, \bibinfo{number}{1} (\bibinfo{year}{2011}), \bibinfo{pages}{93--104}.
\newblock


\bibitem[Hofer et~al\mbox{.}(2009)]%
        {hofer2009sloth}
\bibfield{author}{\bibinfo{person}{Wanja Hofer}, \bibinfo{person}{Daniel Lohmann}, \bibinfo{person}{Fabian Scheler}, {and} \bibinfo{person}{Wolfgang Schr{\"o}der-Preikschat}.} \bibinfo{year}{2009}\natexlab{}.
\newblock \showarticletitle{Sloth: Threads as interrupts}. In \bibinfo{booktitle}{\emph{2009 30th IEEE Real-Time Systems Symposium}}. IEEE, \bibinfo{pages}{204--213}.
\newblock


\bibitem[Farzan et~al\mbox{.}(2009)]%
        {farzan2009meta}
\bibfield{author}{\bibinfo{person}{Azadeh Farzan}, \bibinfo{person}{P Madhusudan}, {and} \bibinfo{person}{Francesco Sorrentino}.} \bibinfo{year}{2009}\natexlab{}.
\newblock \showarticletitle{Meta-analysis for atomicity violations under nested locking}. In \bibinfo{booktitle}{\emph{International Conference on Computer Aided Verification}}. Springer, \bibinfo{pages}{248--262}.
\newblock


\bibitem[Jaulin(2000)]%
        {jaulin2000interval}
\bibfield{author}{\bibinfo{person}{Luc Jaulin}.} \bibinfo{year}{2000}\natexlab{}.
\newblock \showarticletitle{Interval constraint propagation with application to bounded-error estimation}.
\newblock \bibinfo{journal}{\emph{Automatica}} \bibinfo{volume}{36}, \bibinfo{number}{10} (\bibinfo{year}{2000}), \bibinfo{pages}{1547--1552}.
\newblock


\bibitem[Cavus et~al\mbox{.}(2020)]%
        {10.1145/3374216}
\bibfield{author}{\bibinfo{person}{Mustafa Cavus}, \bibinfo{person}{Resit Sendag}, {and} \bibinfo{person}{Joshua~J. Yi}.} \bibinfo{year}{2020}\natexlab{}.
\newblock \showarticletitle{Informed Prefetching for Indirect Memory Accesses}.
\newblock \bibinfo{journal}{\emph{ACM Trans. Archit. Code Optim.}} \bibinfo{volume}{17}, \bibinfo{number}{1}, Article \bibinfo{articleno}{4} (\bibinfo{date}{March} \bibinfo{year}{2020}), \bibinfo{numpages}{29}~pages.
\newblock
\showISSN{1544-3566}
\urldef\tempurl%
\url{https://doi.org/10.1145/3374216}
\showDOI{\tempurl}


\bibitem[Barai et~al\mbox{.}(2024)]%
        {10.1145/3631882.3631885}
\bibfield{author}{\bibinfo{person}{Atanu Barai}, \bibinfo{person}{Nandakishore Santhi}, \bibinfo{person}{Abdur Razzak}, \bibinfo{person}{Stephan Eidenbenz}, {and} \bibinfo{person}{Abdel-Hameed~A. Badawy}.} \bibinfo{year}{2024}\natexlab{}.
\newblock \showarticletitle{LLVM Static Analysis for Program Characterization and Memory Reuse Profile Estimation}. In \bibinfo{booktitle}{\emph{Proceedings of the International Symposium on Memory Systems}} (Alexandria, VA, USA) \emph{(\bibinfo{series}{MEMSYS '23})}. \bibinfo{publisher}{Association for Computing Machinery}, \bibinfo{address}{New York, NY, USA}, Article \bibinfo{articleno}{3}, \bibinfo{numpages}{6}~pages.
\newblock
\showISBNx{9798400716447}
\urldef\tempurl%
\url{https://doi.org/10.1145/3631882.3631885}
\showDOI{\tempurl}


\bibitem[Wei et~al\mbox{.}(2022)]%
        {wei2022chain}
\bibfield{author}{\bibinfo{person}{Jason Wei}, \bibinfo{person}{Xuezhi Wang}, \bibinfo{person}{Dale Schuurmans}, \bibinfo{person}{Maarten Bosma}, \bibinfo{person}{Fei Xia}, \bibinfo{person}{Ed Chi}, \bibinfo{person}{Quoc~V Le}, \bibinfo{person}{Denny Zhou}, {et~al\mbox{.}}} \bibinfo{year}{2022}\natexlab{}.
\newblock \showarticletitle{Chain-of-thought prompting elicits reasoning in large language models}.
\newblock \bibinfo{journal}{\emph{Advances in neural information processing systems}}  \bibinfo{volume}{35} (\bibinfo{year}{2022}), \bibinfo{pages}{24824--24837}.
\newblock


\bibitem[Cla(2020)]%
        {Clang}
 \bibinfo{year}{2020}\natexlab{}.
\newblock \bibinfo{title}{Clang-10.0}.
\newblock \bibinfo{howpublished}{\url{https://clang.llvm.org/}}.
\newblock


\bibitem[LLV(2020)]%
        {LLVM}
 \bibinfo{year}{2020}\natexlab{}.
\newblock \bibinfo{title}{LLVM-10.0}.
\newblock \bibinfo{howpublished}{\url{https://llvm.org/}}.
\newblock


\bibitem[Anthropic(2024)]%
        {claude-4}
\bibfield{author}{\bibinfo{person}{Anthropic}.} \bibinfo{year}{2024}\natexlab{}.
\newblock \bibinfo{title}{Claude-Sonnet-4}.
\newblock \bibinfo{howpublished}{\url{https://www.anthropic.com/claude/sonnet}}.
\newblock


\bibitem[Wang et~al\mbox{.}(2023)]%
        {wang2023boosting}
\bibfield{author}{\bibinfo{person}{Chong Wang}, \bibinfo{person}{Jianan Liu}, \bibinfo{person}{Xin Peng}, \bibinfo{person}{Yang Liu}, {and} \bibinfo{person}{Yiling Lou}.} \bibinfo{year}{2023}\natexlab{}.
\newblock \showarticletitle{Boosting static resource leak detection via llm-based resource-oriented intention inference}.
\newblock \bibinfo{journal}{\emph{CoRR}} (\bibinfo{year}{2023}).
\newblock


\bibitem[Anthropic(2024)]%
        {claude-3.5}
\bibfield{author}{\bibinfo{person}{Anthropic}.} \bibinfo{year}{2024}\natexlab{}.
\newblock \bibinfo{title}{Claude-3.5-Sonnet}.
\newblock \bibinfo{howpublished}{\url{https://www.anthropic.com/claude/sonnet}}.
\newblock


\bibitem[Comanici et~al\mbox{.}(2025)]%
        {comanici2025gemini}
\bibfield{author}{\bibinfo{person}{Gheorghe Comanici}, \bibinfo{person}{Eric Bieber}, \bibinfo{person}{Mike Schaekermann}, \bibinfo{person}{Ice Pasupat}, \bibinfo{person}{Noveen Sachdeva}, \bibinfo{person}{Inderjit Dhillon}, \bibinfo{person}{Marcel Blistein}, \bibinfo{person}{Ori Ram}, \bibinfo{person}{Dan Zhang}, \bibinfo{person}{Evan Rosen}, {et~al\mbox{.}}} \bibinfo{year}{2025}\natexlab{}.
\newblock \showarticletitle{Gemini 2.5: Pushing the frontier with advanced reasoning, multimodality, long context, and next generation agentic capabilities}.
\newblock \bibinfo{journal}{\emph{arXiv preprint arXiv:2507.06261}} (\bibinfo{year}{2025}).
\newblock


\bibitem[Team(2024)]%
        {qwen25}
\bibfield{author}{\bibinfo{person}{Qwen Team}.} \bibinfo{year}{2024}\natexlab{}.
\newblock \showarticletitle{Qwen2.5 technical report}.
\newblock \bibinfo{journal}{\emph{arXiv preprint arXiv:2412.15115}} (\bibinfo{year}{2024}).
\newblock


\bibitem[Team et~al\mbox{.}(2025)]%
        {team2025kimi}
\bibfield{author}{\bibinfo{person}{Kimi Team}, \bibinfo{person}{Yifan Bai}, \bibinfo{person}{Yiping Bao}, \bibinfo{person}{Guanduo Chen}, \bibinfo{person}{Jiahao Chen}, \bibinfo{person}{Ningxin Chen}, \bibinfo{person}{Ruijue Chen}, \bibinfo{person}{Yanru Chen}, \bibinfo{person}{Yuankun Chen}, \bibinfo{person}{Yutian Chen}, {et~al\mbox{.}}} \bibinfo{year}{2025}\natexlab{}.
\newblock \showarticletitle{Kimi k2: Open agentic intelligence}.
\newblock \bibinfo{journal}{\emph{arXiv preprint arXiv:2507.20534}} (\bibinfo{year}{2025}).
\newblock


\bibitem[Liu et~al\mbox{.}(2024)]%
        {liu2024deepseek}
\bibfield{author}{\bibinfo{person}{Aixin Liu}, \bibinfo{person}{Bei Feng}, \bibinfo{person}{Bing Xue}, \bibinfo{person}{Bingxuan Wang}, \bibinfo{person}{Bochao Wu}, \bibinfo{person}{Chengda Lu}, \bibinfo{person}{Chenggang Zhao}, \bibinfo{person}{Chengqi Deng}, \bibinfo{person}{Chenyu Zhang}, \bibinfo{person}{Chong Ruan}, {et~al\mbox{.}}} \bibinfo{year}{2024}\natexlab{}.
\newblock \showarticletitle{Deepseek-v3 technical report}.
\newblock \bibinfo{journal}{\emph{arXiv preprint arXiv:2412.19437}} (\bibinfo{year}{2024}).
\newblock


\bibitem[Team et~al\mbox{.}(2025)]%
        {longcatflashtechnicalreport}
\bibfield{author}{\bibinfo{person}{Meituan~LongCat Team}, \bibinfo{person}{Bayan}, \bibinfo{person}{Bei Li}, \bibinfo{person}{Bingye Lei}, \bibinfo{person}{Bo Wang}, \bibinfo{person}{Bolin Rong}, \bibinfo{person}{Chao Wang}, \bibinfo{person}{Chao Zhang}, \bibinfo{person}{Chen Gao}, \bibinfo{person}{Chen Zhang}, \bibinfo{person}{Cheng Sun}, \bibinfo{person}{Chengcheng Han}, \bibinfo{person}{Chenguang Xi}, \bibinfo{person}{Chi Zhang}, \bibinfo{person}{Chong Peng}, \bibinfo{person}{Chuan Qin}, \bibinfo{person}{Chuyu Zhang}, \bibinfo{person}{Cong Chen}, \bibinfo{person}{Congkui Wang}, \bibinfo{person}{Dan Ma}, \bibinfo{person}{Daoru Pan}, \bibinfo{person}{Defei Bu}, \bibinfo{person}{Dengchang Zhao}, \bibinfo{person}{Deyang Kong}, \bibinfo{person}{Dishan Liu}, \bibinfo{person}{Feiye Huo}, \bibinfo{person}{Fengcun Li}, \bibinfo{person}{Fubao Zhang}, \bibinfo{person}{Gan Dong}, \bibinfo{person}{Gang Liu}, \bibinfo{person}{Gang Xu}, \bibinfo{person}{Ge Li}, \bibinfo{person}{Guoqiang Tan}, \bibinfo{person}{Guoyuan
  Lin}, \bibinfo{person}{Haihang Jing}, \bibinfo{person}{Haomin Fu}, \bibinfo{person}{Haonan Yan}, \bibinfo{person}{Haoxing Wen}, \bibinfo{person}{Haozhe Zhao}, \bibinfo{person}{Hong Liu}, \bibinfo{person}{Hongmei Shi}, \bibinfo{person}{Hongyan Hao}, \bibinfo{person}{Hongyin Tang}, \bibinfo{person}{Huantian Lv}, \bibinfo{person}{Hui Su}, \bibinfo{person}{Jiacheng Li}, \bibinfo{person}{Jiahao Liu}, \bibinfo{person}{Jiahuan Li}, \bibinfo{person}{Jiajun Yang}, \bibinfo{person}{Jiaming Wang}, \bibinfo{person}{Jian Yang}, \bibinfo{person}{Jianchao Tan}, \bibinfo{person}{Jiaqi Sun}, \bibinfo{person}{Jiaqi Zhang}, \bibinfo{person}{Jiawei Fu}, \bibinfo{person}{Jiawei Yang}, \bibinfo{person}{Jiaxi Hu}, \bibinfo{person}{Jiayu Qin}, \bibinfo{person}{Jingang Wang}, \bibinfo{person}{Jiyuan He}, \bibinfo{person}{Jun Kuang}, \bibinfo{person}{Junhui Mei}, \bibinfo{person}{Kai Liang}, \bibinfo{person}{Ke He}, \bibinfo{person}{Kefeng Zhang}, \bibinfo{person}{Keheng Wang}, \bibinfo{person}{Keqing He}, \bibinfo{person}{Liang
  Gao}, \bibinfo{person}{Liang Shi}, \bibinfo{person}{Lianhui Ma}, \bibinfo{person}{Lin Qiu}, \bibinfo{person}{Lingbin Kong}, \bibinfo{person}{Lingtong Si}, \bibinfo{person}{Linkun Lyu}, \bibinfo{person}{Linsen Guo}, \bibinfo{person}{Liqi Yang}, \bibinfo{person}{Lizhi Yan}, \bibinfo{person}{Mai Xia}, \bibinfo{person}{Man Gao}, \bibinfo{person}{Manyuan Zhang}, \bibinfo{person}{Meng Zhou}, \bibinfo{person}{Mengxia Shen}, \bibinfo{person}{Mingxiang Tuo}, \bibinfo{person}{Mingyang Zhu}, \bibinfo{person}{Peiguang Li}, \bibinfo{person}{Peng Pei}, \bibinfo{person}{Peng Zhao}, \bibinfo{person}{Pengcheng Jia}, \bibinfo{person}{Pingwei Sun}, \bibinfo{person}{Qi Gu}, \bibinfo{person}{Qianyun Li}, \bibinfo{person}{Qingyuan Li}, \bibinfo{person}{Qiong Huang}, \bibinfo{person}{Qiyuan Duan}, \bibinfo{person}{Ran Meng}, \bibinfo{person}{Rongxiang Weng}, \bibinfo{person}{Ruichen Shao}, \bibinfo{person}{Rumei Li}, \bibinfo{person}{Shizhe Wu}, \bibinfo{person}{Shuai Liang}, \bibinfo{person}{Shuo Wang}, \bibinfo{person}{Suogui
  Dang}, \bibinfo{person}{Tao Fang}, \bibinfo{person}{Tao Li}, \bibinfo{person}{Tefeng Chen}, \bibinfo{person}{Tianhao Bai}, \bibinfo{person}{Tianhao Zhou}, \bibinfo{person}{Tingwen Xie}, \bibinfo{person}{Wei He}, \bibinfo{person}{Wei Huang}, \bibinfo{person}{Wei Liu}, \bibinfo{person}{Wei Shi}, \bibinfo{person}{Wei Wang}, \bibinfo{person}{Wei Wu}, \bibinfo{person}{Weikang Zhao}, \bibinfo{person}{Wen Zan}, \bibinfo{person}{Wenjie Shi}, \bibinfo{person}{Xi Nan}, \bibinfo{person}{Xi Su}, \bibinfo{person}{Xiang Li}, \bibinfo{person}{Xiang Mei}, \bibinfo{person}{Xiangyang Ji}, \bibinfo{person}{Xiangyu Xi}, \bibinfo{person}{Xiangzhou Huang}, \bibinfo{person}{Xianpeng Li}, \bibinfo{person}{Xiao Fu}, \bibinfo{person}{Xiao Liu}, \bibinfo{person}{Xiao Wei}, \bibinfo{person}{Xiaodong Cai}, \bibinfo{person}{Xiaolong Chen}, \bibinfo{person}{Xiaoqing Liu}, \bibinfo{person}{Xiaotong Li}, \bibinfo{person}{Xiaowei Shi}, \bibinfo{person}{Xiaoyu Li}, \bibinfo{person}{Xili Wang}, \bibinfo{person}{Xin Chen},
  \bibinfo{person}{Xing Hu}, \bibinfo{person}{Xingyu Miao}, \bibinfo{person}{Xinyan He}, \bibinfo{person}{Xuemiao Zhang}, \bibinfo{person}{Xueyuan Hao}, \bibinfo{person}{Xuezhi Cao}, \bibinfo{person}{Xunliang Cai}, \bibinfo{person}{Xurui Yang}, \bibinfo{person}{Yan Feng}, \bibinfo{person}{Yang Bai}, \bibinfo{person}{Yang Chen}, \bibinfo{person}{Yang Yang}, \bibinfo{person}{Yaqi Huo}, \bibinfo{person}{Yerui Sun}, \bibinfo{person}{Yifan Lu}, \bibinfo{person}{Yifan Zhang}, \bibinfo{person}{Yipeng Zang}, \bibinfo{person}{Yitao Zhai}, \bibinfo{person}{Yiyang Li}, \bibinfo{person}{Yongjing Yin}, \bibinfo{person}{Yongkang Lv}, \bibinfo{person}{Yongwei Zhou}, \bibinfo{person}{Yu Yang}, \bibinfo{person}{Yuchen Xie}, \bibinfo{person}{Yueqing Sun}, \bibinfo{person}{Yuewen Zheng}, \bibinfo{person}{Yuhua Wei}, \bibinfo{person}{Yulei Qian}, \bibinfo{person}{Yunfan Liang}, \bibinfo{person}{Yunfang Tai}, \bibinfo{person}{Yunke Zhao}, \bibinfo{person}{Zeyang Yu}, \bibinfo{person}{Zhao Zhang}, \bibinfo{person}{Zhaohua Yang},
  \bibinfo{person}{Zhenchao Zhang}, \bibinfo{person}{Zhikang Xia}, \bibinfo{person}{Zhiye Zou}, \bibinfo{person}{Zhizhao Zeng}, \bibinfo{person}{Zhongda Su}, \bibinfo{person}{Zhuofan Chen}, \bibinfo{person}{Zijian Zhang}, \bibinfo{person}{Ziwen Wang}, \bibinfo{person}{Zixu Jiang}, \bibinfo{person}{Zizhe Zhao}, \bibinfo{person}{Zongyu Wang}, {and} \bibinfo{person}{Zunhai Su}.} \bibinfo{year}{2025}\natexlab{}.
\newblock \bibinfo{title}{LongCat-Flash Technical Report}.
\newblock
\newblock
\showeprint[arxiv]{2509.01322}~[cs.CL]
\urldef\tempurl%
\url{https://arxiv.org/abs/2509.01322}
\showURL{%
\tempurl}


\bibitem[Aigcbest(2025)]%
        {aigcbest}
\bibfield{author}{\bibinfo{person}{Aigcbest}.} \bibinfo{year}{2025}\natexlab{}.
\newblock \bibinfo{title}{Aigcbest}.
\newblock \bibinfo{howpublished}{\url{https://api.aigcbest.top/}}.
\newblock


\bibitem[Chen et~al\mbox{.}(2023)]%
        {DRB-LLM}
\bibfield{author}{\bibinfo{person}{Le Chen}, \bibinfo{person}{Xianzhong Ding}, \bibinfo{person}{Murali Emani}, \bibinfo{person}{Tristan Vanderbruggen}, \bibinfo{person}{Pei-Hung Lin}, {and} \bibinfo{person}{Chunhua Liao}.} \bibinfo{year}{2023}\natexlab{}.
\newblock \showarticletitle{Data Race Detection Using Large Language Models} \emph{(\bibinfo{series}{SC-W '23})}. \bibinfo{publisher}{Association for Computing Machinery}, \bibinfo{address}{New York, NY, USA}, \bibinfo{pages}{215--223}.
\newblock
\showISBNx{9798400707858}
\urldef\tempurl%
\url{https://doi.org/10.1145/3624062.3624088}
\showDOI{\tempurl}


\bibitem[Liao et~al\mbox{.}(2017)]%
        {DataRaceBench}
\bibfield{author}{\bibinfo{person}{Chunhua Liao}, \bibinfo{person}{Pei-Hung Lin}, \bibinfo{person}{Joshua Asplund}, \bibinfo{person}{Markus Schordan}, {and} \bibinfo{person}{Ian Karlin}.} \bibinfo{year}{2017}\natexlab{}.
\newblock \showarticletitle{DataRaceBench: a benchmark suite for systematic evaluation of data race detection tools}. In \bibinfo{booktitle}{\emph{Proceedings of the International Conference for High Performance Computing, Networking, Storage and Analysis}} (Denver, Colorado) \emph{(\bibinfo{series}{SC '17})}. \bibinfo{publisher}{Association for Computing Machinery}, \bibinfo{address}{New York, NY, USA}, Article \bibinfo{articleno}{11}, \bibinfo{numpages}{14}~pages.
\newblock
\showISBNx{9781450351140}
\urldef\tempurl%
\url{https://doi.org/10.1145/3126908.3126958}
\showDOI{\tempurl}


\bibitem[Wang et~al\mbox{.}(2022)]%
        {wang2022specchecker}
\bibfield{author}{\bibinfo{person}{Boxiang Wang}, \bibinfo{person}{Rui Chen}, \bibinfo{person}{Chao Li}, \bibinfo{person}{Tingting Yu}, \bibinfo{person}{Dongdong Gao}, {and} \bibinfo{person}{Mengfei Yang}.} \bibinfo{year}{2022}\natexlab{}.
\newblock \showarticletitle{SpecChecker-ISA: a data sharing analyzer for interrupt-driven embedded software}. In \bibinfo{booktitle}{\emph{Proceedings of the 31st ACM SIGSOFT International Symposium on Software Testing and Analysis}}. \bibinfo{pages}{801--804}.
\newblock


\bibitem[Lu et~al\mbox{.}(2012)]%
        {5740930}
\bibfield{author}{\bibinfo{person}{Shan Lu}, \bibinfo{person}{Soyeon Park}, {and} \bibinfo{person}{Yuanyuan Zhou}.} \bibinfo{year}{2012}\natexlab{}.
\newblock \showarticletitle{Finding Atomicity-Violation Bugs through Unserializable Interleaving Testing}.
\newblock \bibinfo{journal}{\emph{IEEE Transactions on Software Engineering}} \bibinfo{volume}{38}, \bibinfo{number}{4} (\bibinfo{year}{2012}), \bibinfo{pages}{844--860}.
\newblock
\urldef\tempurl%
\url{https://doi.org/10.1109/TSE.2011.35}
\showDOI{\tempurl}


\bibitem[Wang et~al\mbox{.}(2025)]%
        {wang2025contemporary}
\bibfield{author}{\bibinfo{person}{Jiayimei Wang}, \bibinfo{person}{Tao Ni}, \bibinfo{person}{Wei-Bin Lee}, {and} \bibinfo{person}{Qingchuan Zhao}.} \bibinfo{year}{2025}\natexlab{}.
\newblock \showarticletitle{A Contemporary Survey of Large Language Model Assisted Program Analysis}.
\newblock \bibinfo{journal}{\emph{arXiv preprint arXiv:2502.18474}} (\bibinfo{year}{2025}).
\newblock


\bibitem[Chapman et~al\mbox{.}(2024)]%
        {10.1145/3652588.3663317}
\bibfield{author}{\bibinfo{person}{Patrick~J. Chapman}, \bibinfo{person}{Cindy Rubio-Gonz\'{a}lez}, {and} \bibinfo{person}{Aditya~V. Thakur}.} \bibinfo{year}{2024}\natexlab{}.
\newblock \showarticletitle{Interleaving Static Analysis and LLM Prompting}. In \bibinfo{booktitle}{\emph{Proceedings of the 13th ACM SIGPLAN International Workshop on the State Of the Art in Program Analysis}} (Copenhagen, Denmark) \emph{(\bibinfo{series}{SOAP 2024})}. \bibinfo{pages}{9--17}.
\newblock
\showISBNx{9798400706219}
\urldef\tempurl%
\url{https://doi.org/10.1145/3652588.3663317}
\showDOI{\tempurl}


\bibitem[Hao et~al\mbox{.}(2023)]%
        {hao2023evpromptinglargelanguage}
\bibfield{author}{\bibinfo{person}{Yu Hao}, \bibinfo{person}{Weiteng Chen}, \bibinfo{person}{Ziqiao Zhou}, {and} \bibinfo{person}{Weidong Cui}.} \bibinfo{year}{2023}\natexlab{}.
\newblock \bibinfo{title}{E\&V: Prompting Large Language Models to Perform Static Analysis by Pseudo-code Execution and Verification}.
\newblock
\newblock
\showeprint[arxiv]{2312.08477}
\urldef\tempurl%
\url{https://arxiv.org/abs/2312.08477}
\showURL{%
\tempurl}


\bibitem[Li et~al\mbox{.}(2024)]%
        {li2024enhancing}
\bibfield{author}{\bibinfo{person}{Haonan Li}, \bibinfo{person}{Yu Hao}, \bibinfo{person}{Yizhuo Zhai}, {and} \bibinfo{person}{Zhiyun Qian}.} \bibinfo{year}{2024}\natexlab{}.
\newblock \showarticletitle{Enhancing Static Analysis for Practical Bug Detection: An LLM-Integrated Approach}.
\newblock \bibinfo{journal}{\emph{Proceedings of the ACM on Programming Languages}} \bibinfo{volume}{8}, \bibinfo{number}{OOPSLA1} (\bibinfo{year}{2024}), \bibinfo{pages}{474--499}.
\newblock


\bibitem[Wang et~al\mbox{.}(2024)]%
        {wang2024llmdfa}
\bibfield{author}{\bibinfo{person}{Chengpeng Wang}, \bibinfo{person}{Wuqi Zhang}, \bibinfo{person}{Zian Su}, \bibinfo{person}{Xiangzhe Xu}, \bibinfo{person}{Xiaoheng Xie}, {and} \bibinfo{person}{Xiangyu Zhang}.} \bibinfo{year}{2024}\natexlab{}.
\newblock \showarticletitle{{LLMDFA}: Analyzing Dataflow in Code with Large Language Models}. In \bibinfo{booktitle}{\emph{The Thirty-eighth Annual Conference on Neural Information Processing Systems}}.
\newblock
\urldef\tempurl%
\url{https://openreview.net/forum?id=QZ2d8E8Whu}
\showURL{%
\tempurl}


\bibitem[Liu et~al\mbox{.}(2024)]%
        {liu2024large}
\bibfield{author}{\bibinfo{person}{Junwei Liu}, \bibinfo{person}{Kaixin Wang}, \bibinfo{person}{Yixuan Chen}, \bibinfo{person}{Xin Peng}, \bibinfo{person}{Zhenpeng Chen}, \bibinfo{person}{Lingming Zhang}, {and} \bibinfo{person}{Yiling Lou}.} \bibinfo{year}{2024}\natexlab{}.
\newblock \showarticletitle{Large language model-based agents for software engineering: A survey}.
\newblock \bibinfo{journal}{\emph{arXiv preprint arXiv:2409.02977}} (\bibinfo{year}{2024}).
\newblock


\bibitem[Du et~al\mbox{.}(2024)]%
        {du2024vul}
\bibfield{author}{\bibinfo{person}{Xueying Du}, \bibinfo{person}{Geng Zheng}, \bibinfo{person}{Kaixin Wang}, \bibinfo{person}{Jiayi Feng}, \bibinfo{person}{Wentai Deng}, \bibinfo{person}{Mingwei Liu}, \bibinfo{person}{Bihuan Chen}, \bibinfo{person}{Xin Peng}, \bibinfo{person}{Tao Ma}, {and} \bibinfo{person}{Yiling Lou}.} \bibinfo{year}{2024}\natexlab{}.
\newblock \showarticletitle{Vul-RAG: Enhancing LLM-based Vulnerability Detection via Knowledge-level RAG}.
\newblock \bibinfo{journal}{\emph{arXiv preprint arXiv:2406.11147}} (\bibinfo{year}{2024}).
\newblock


\bibitem[Zhang et~al\mbox{.}(2024)]%
        {zhang2024codeagent}
\bibfield{author}{\bibinfo{person}{Kechi Zhang}, \bibinfo{person}{Jia Li}, \bibinfo{person}{Ge Li}, \bibinfo{person}{Xianjie Shi}, {and} \bibinfo{person}{Zhi Jin}.} \bibinfo{year}{2024}\natexlab{}.
\newblock \showarticletitle{Codeagent: Enhancing code generation with tool-integrated agent systems for real-world repo-level coding challenges}.
\newblock \bibinfo{journal}{\emph{arXiv preprint arXiv:2401.07339}} (\bibinfo{year}{2024}).
\newblock


\bibitem[Chen et~al\mbox{.}(2024)]%
        {chen2024coder}
\bibfield{author}{\bibinfo{person}{Dong Chen}, \bibinfo{person}{Shaoxin Lin}, \bibinfo{person}{Muhan Zeng}, \bibinfo{person}{Daoguang Zan}, \bibinfo{person}{Jian-Gang Wang}, \bibinfo{person}{Anton Cheshkov}, \bibinfo{person}{Jun Sun}, \bibinfo{person}{Hao Yu}, \bibinfo{person}{Guoliang Dong}, \bibinfo{person}{Artem Aliev}, {et~al\mbox{.}}} \bibinfo{year}{2024}\natexlab{}.
\newblock \showarticletitle{CodeR: Issue Resolving with Multi-Agent and Task Graphs}.
\newblock \bibinfo{journal}{\emph{arXiv preprint arXiv:2406.01304}} (\bibinfo{year}{2024}).
\newblock


\bibitem[Liu et~al\mbox{.}(2024)]%
        {liu2024marscode}
\bibfield{author}{\bibinfo{person}{Yizhou Liu}, \bibinfo{person}{Pengfei Gao}, \bibinfo{person}{Xinchen Wang}, \bibinfo{person}{Jie Liu}, \bibinfo{person}{Yexuan Shi}, \bibinfo{person}{Zhao Zhang}, {and} \bibinfo{person}{Chao Peng}.} \bibinfo{year}{2024}\natexlab{}.
\newblock \showarticletitle{MarsCode Agent: AI-native Automated Bug Fixing}.
\newblock \bibinfo{journal}{\emph{arXiv preprint arXiv:2409.00899}} (\bibinfo{year}{2024}).
\newblock


\bibitem[Mao et~al\mbox{.}(2024)]%
        {mao2024multi}
\bibfield{author}{\bibinfo{person}{Zhenyy Mao}, \bibinfo{person}{Jialong Li}, \bibinfo{person}{Dongming Jin}, \bibinfo{person}{Munan Li}, {and} \bibinfo{person}{Kenji Tei}.} \bibinfo{year}{2024}\natexlab{}.
\newblock \showarticletitle{Multi-role consensus through llms discussions for vulnerability detection}. In \bibinfo{booktitle}{\emph{2024 IEEE 24th International Conference on Software Quality, Reliability, and Security Companion (QRS-C)}}. IEEE, \bibinfo{pages}{1318--1319}.
\newblock


\bibitem[Islam et~al\mbox{.}(2024)]%
        {islam2024mapcoder}
\bibfield{author}{\bibinfo{person}{Md~Ashraful Islam}, \bibinfo{person}{Mohammed~Eunus Ali}, {and} \bibinfo{person}{Md~Rizwan Parvez}.} \bibinfo{year}{2024}\natexlab{}.
\newblock \showarticletitle{MapCoder: Multi-Agent Code Generation for Competitive Problem Solving}.
\newblock \bibinfo{journal}{\emph{arXiv preprint arXiv:2405.11403}} (\bibinfo{year}{2024}).
\newblock


\bibitem[Nguyen et~al\mbox{.}(2024)]%
        {nguyen2024agilecoder}
\bibfield{author}{\bibinfo{person}{Minh~Huynh Nguyen}, \bibinfo{person}{Thang~Phan Chau}, \bibinfo{person}{Phong~X Nguyen}, {and} \bibinfo{person}{Nghi~DQ Bui}.} \bibinfo{year}{2024}\natexlab{}.
\newblock \showarticletitle{AgileCoder: Dynamic Collaborative Agents for Software Development based on Agile Methodology}.
\newblock \bibinfo{journal}{\emph{arXiv preprint arXiv:2406.11912}} (\bibinfo{year}{2024}).
\newblock


\bibitem[Zhang et~al\mbox{.}(2024)]%
        {zhang2024autocoderover}
\bibfield{author}{\bibinfo{person}{Yuntong Zhang}, \bibinfo{person}{Haifeng Ruan}, \bibinfo{person}{Zhiyu Fan}, {and} \bibinfo{person}{Abhik Roychoudhury}.} \bibinfo{year}{2024}\natexlab{}.
\newblock \showarticletitle{Autocoderover: Autonomous program improvement}. In \bibinfo{booktitle}{\emph{Proceedings of the 33rd ACM SIGSOFT International Symposium on Software Testing and Analysis}}. \bibinfo{pages}{1592--1604}.
\newblock


\bibitem[Qian et~al\mbox{.}(2024)]%
        {qian2024scaling}
\bibfield{author}{\bibinfo{person}{Chen Qian}, \bibinfo{person}{Zihao Xie}, \bibinfo{person}{Yifei Wang}, \bibinfo{person}{Wei Liu}, \bibinfo{person}{Yufan Dang}, \bibinfo{person}{Zhuoyun Du}, \bibinfo{person}{Weize Chen}, \bibinfo{person}{Cheng Yang}, \bibinfo{person}{Zhiyuan Liu}, {and} \bibinfo{person}{Maosong Sun}.} \bibinfo{year}{2024}\natexlab{}.
\newblock \showarticletitle{Scaling Large-Language-Model-based Multi-Agent Collaboration}.
\newblock \bibinfo{journal}{\emph{arXiv preprint arXiv:2406.07155}} (\bibinfo{year}{2024}).
\newblock


\end{thebibliography}
\end{document}